\newif\ifmulticol	\multicoltrue
\newif\ifshowgit	\showgittrue		
\newif\ifgitlocal	\gitlocalfalse		
\newif\ifbiblatex	\biblatexfalse		
\newif\ifbibnum		\bibnumtrue 		
\newif\ifbibsort	\bibsortfalse		
\newif\iflineno		\linenofalse
\newif\iftoc		\toctrue

\newif\iflucida		\lucidatrue
\newif\ifcm			\cmfalse
\newif\ifcharter	\charterfalse		

\newcommand*{\secnumdepth}{0}			
\newcommand*{\tocdepth}{2}				
\newcommand*{\reftitle}{References}		
\newcommand*{\mytitleskip}{-0.2in}		

\multicoltrue\showgittrue\gitlocaltrue\lucidafalse\cmtrue

\documentclass[11pt,twocolumn]{article}

\newcommand*{\pdfauthor}{S.~A.~Frank}
\newcommand*{\pdftitle}{Natural selection at multiple scales}

\usepackage{afterpage}
\usepackage{blkarray}

\input mls.sty

\def\qbar{\bar{q}}

\newcommand{\Cm}{C_m}
\newcommand{\Bm}{B_m}
\newcommand{\cov}{{\hbox{\rm cov}}}
\newcommand{\var}{{\hbox{\rm var}}}
\newcommand{\byx}{\Gb_{yx}}

\newcommand*{\Ga}{\alpha}
\newcommand*{\Gb}{\beta}
\newcommand*{\Gd}{\delta}

\newcommand*{\Gg}{\gamma}

\newcommand*{\Gk}{\kappa}
\newcommand*{\Gl}{\lambda}

\newcommand*{\Gm}{\mu}

\newcommand*{\Gr}{\rho}

\newcommand*{\Gth}{\theta}
\newcommand*{\Gf}{\phi}

\usepackage{bm}
\newcommand{\bmr}[1]{\bm{\mathrm{#1}}}

\DeclarePairedDelimiter\abs{\lvert}{\rvert}
\DeclarePairedDelimiter\norm{\lVert}{\rVert}
\DeclarePairedDelimiter\angb{\langle}{\rangle}
\DeclarePairedDelimiter\lrb{\lbrack}{\rbrack}
\DeclarePairedDelimiter\lr{\lparen}{\rparen}
\DeclarePairedDelimiter\lrbr{\lbrace}{\rbrace}

\makeatletter
\let\oldabs\abs \def\abs{\@ifstar{\oldabs}{\oldabs*}}
\let\oldnorm\norm \def\norm{\@ifstar{\oldnorm}{\oldnorm*}}
\let\oldangb\angb \def\angb{\@ifstar{\oldangb}{\oldangb*}}
\let\oldlrb\lrb \def\lrb{\@ifstar{\oldlrb}{\oldlrb*}}
\let\oldlr\lr \def\lr{\@ifstar{\oldlr}{\oldlr*}}
\let\oldlrbr\lrbr \def\lrbr{\@ifstar{\oldlrbr}{\oldlrbr*}}
\makeatother

\newcommand*{\dd}{\textrm{d}}

\newcommand*{\Eq}[1]{eqn~\ref{eq:#1}}
\newcommand*{\Eqq}[1]{eqns~\ref{eq:#1}}

\newcommand*{\dovr}[2]{\frac{\dd #1}{\dd #2}}

\newcommand*{\prt}{\partial}
\newcommand*{\povr}[2]{\frac{\prt #1}{\prt #2}}

\newcommand*{\Figure}[1]{Figure~\ref{fig:#1}}
\newcommand*{\Fig}[1]{Fig.~\ref{fig:#1}}

\newcommand*{\boldrule}{\hrule height 1.2pt}
\newcommand*{\noterule}{\medskip\boldrule\medskip}	

\hypersetup{linkcolor={Maroon4}, urlcolor={Maroon4}}

\newcommand*{\mytitle}{\pdftitle}

\newcommand*{\myauthor}{Steven A.~Frank}
\newcommand*{\myaddress}{Department of Ecology and Evolutionary Biology, University of California, Irvine, CA 92697--2525, USA; email: safrank@uci.edu}

\newcommand*{\mykeywords}{Evolutionary theory, mathematical models, group selection, kin selection, spatial processes, statics vs dynamics}

\setabstract{-0.07}{\iftoc-1.6\else0.0\fi}{%
Natural selection acts on traits at different scales, often with opposing consequences. This article identifies the particular forces that act at each scale and how those forces combine to determine the overall evolutionary outcome. A series of extended models derive from the tragedy of the commons, illustrating opposing forces at different scales. Examples include the primary tension between conflict and cooperation and the evolution of virulence, sex ratios, dispersal, and evolvability. The unified analysis subsumes interactions within and between species by generalizing multitrait interactions. Expanded notions of recombination and cotransmission arise. The core theoretical approach isolates the fundamental forces of selection, including marginal valuation, correlation between interacting entities, and reproductive value. Those fundamental forces act as partial causes that combine at different temporal and spatial scales. Modeling focuses on statics, in the sense of how different forces at various scales tend to oppose each other, ultimately combining to shape traits. That type of static analysis emphasizes explanation rather than the calculation of dynamics. The article builds on the duality between explanation versus calculation in terms of statics versus dynamics. The literature often poses that duality as a controversy, whereas this article develops the pair as complementary tools that together provide deeper understanding. Along the way, the unified approach clarifies the subtle distinctions between kin selection, multilevel selection, and inclusive fitness, subsuming these topics into the broader perspectives of fundamental forces and multiple scales.
}

\begin{document}

\mymaketitle

\iftoc\mytoc{-24pt}{\newpage}\fi

\section{Introduction}

Natural selection often works in opposing ways. Mutation continuously degrades traits but once in a while triggers adaptation to a novel challenge. Frequent competitive gains against neighbors may be offset by the occasional failure of overly competitive groups.

In these examples, selection often pushes one way, occasionally it pushes the other way. Do shorter or longer time scales dominate? When opposing forces happen over different spatial scales, do nearby or distant scales dominate?

These temporal and spatial scales of selection relate to the hierarchical levels of biological organization \autocite{williams66adaptation}. Above the individual, groups may compete for resources. At the species level, rates of extinction and speciation vary.

Selection at each level links to a particular temporal frequency and spatial extent. Multilevel analysis provides a method to study various puzzles \autocite{hamilton75innate}.

Do the female-biased sex ratios often seen in small isolated groups arise by kin selection or group selection? Does species level selection explain a lot of what we see or is it just a weak force that rarely alters pattern? Do synergistic interactions between species create more competitive communities?

These questions concern how selection at various levels and scales ultimately resolves to shape biological traits. In the past, people emphasized alternative perspectives for multiscale problems, suggesting that the different views compete for the best perspective \autocite{maynard-smith76group,wilson83the-group,grafen84natural,queller91group,okasha06evolution,west08social,gardner09capturing,leigh10the-group}.

I show the benefits of combining approaches to get a better picture of the whole. From this broad view, a sharp focus on the opposing forces at different temporal and spatial scales provides a particularly useful way to see recurring aspects in seemingly different problems.

I build my argument through a series of specific models, each designed to highlight the resolution of opposing forces at different scales. For example, competition between nearby individuals favors rapacious traits that degrade the success of the local group, whereas competition between distant groups favors prudent traits that enhance the success of the local group \autocite{wilson75a-theory,leigh77how-does}. Within that context, I clarify subtle aspects of kin selection, group selection, and inclusive fitness.

I extend this common model of spatial scale to include temporal aspects. For example, local competition typically happens more frequently than distant competition. A similar contrast between time scales applies to competition within species versus competition between species \autocite{williams66adaptation}.

Combinations of hierarchical levels, spatial processes, and temporal processes lead to a series of examples, including sex ratios, parasite virulence, dispersal, evolvability, mutation-selection balance, recombination-selection balance, and the synergistic symbioses of integrated communities.

The examples taken together call for a broader conceptual foundation. To develop that foundation, I emphasize common themes and controversies in models of natural selection, including the contrast between equilibrium models and dynamic models and the roles of calculation versus explanation in theoretical analysis.

My examples focus on a small list of fundamental forces of natural selection \autocite{frank98foundations,frank22microbial}. Each force acts as a partial cause in any particular model of natural history. By emphasizing the opposing partial causes that often arise in nature, we gain a clearer understanding of how different scales of selection work together to shape traits.

Such explanation is distinct from the calculation of consequences in more complex models. Calculation allows us to quantify outcomes, whereas explanation provides the conceptual framework to understand why those outcomes occur. A mature theory develops both explanation and calculation.

Overall, this article works toward such a mature theory. I primarily focus on simple models and explanations to balance the well-developed prior work on calculation. Through these models and explanations, this article provides the foundation for understanding how natural selection operates across scales, offering new insights into both well-known and emerging evolutionary puzzles.

\begin{figure*}[t]
\centering
\includegraphics[width=0.60\hsize]{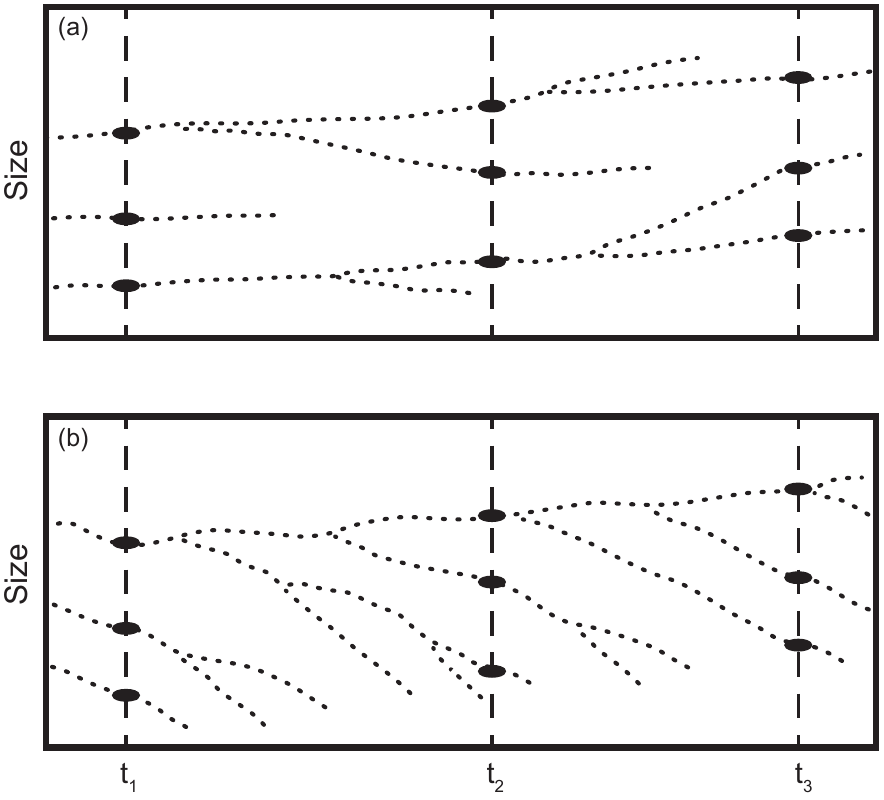}
\vskip0pt
\caption{Alternative mechanisms for an increase in size among a hypothetical group of related species. At the three sampled times, the upper and lower panels show the same increasing distribution of sizes. (a) In this case, size tends to increase within each lineage. The overall increase in size among species is consistent with natural selection favoring larger size within each lineage. (b) Alternatively, size may tend to decrease within each species. However, the largest species survives longer than the others. Overall, the greater success of the largest species relative to the smaller ones outweighs the tendency for size to decrease within most lineages. Selection between species dominates selection between individuals within species. Redrawn from Williams \autocite{williams66adaptation}.
}
\label{fig:williams}
\end{figure*}

\section{Essence}

This section provides a concise overview of the key examples and broader conceptual themes, emphasizing the role of scale. Later sections build on this overview.

\subsection{Opposing levels vs multiple scales}

Imagine a hypothetical pattern for the evolution of horses \autocite{williams66adaptation}. Three samples from the fossil record show that horse size has increased over time. In \Fig{williams}a, individual competition within species favored larger horses over smaller ones. Size tended to increase over time within most species.

Alternatively, in \Fig{williams}b, individual competition typically favored smaller horses within species. That individual level of selection is opposed by the greater survival of the lineage with the largest size. Overall, species-level selection dominates, causing size to increase over time even though individual selection favors smaller horses.

This example highlights opposing selection at two discrete levels. In reality, selection likely works at multiple scales.

Within species, competition occurs frequently between neighbors. Migration and competition over longer spatial scales happen less often. Higher levels such as species births and deaths or the major radiations that create higher taxonomic scales happen with decreasing prevalence.

The small local scales often dominate because they happen much more frequently than the larger scales \autocite{williams66adaptation}. But sometimes the larger scales play an important role.

\subsection{Tragedy of the commons}

Suppose some individuals compete within an isolated patch. Rapacious types extract resources rapidly and inefficiently, outcompete their neighbors, and export more offspring than their local competitors. Rapacity wins at the local scale.

Another patch may be dominated by more prudent types. They use resources more slowly and efficiently. A prudent group ultimately exports more offspring than a rapacious group. At a global scale, prudent patches outcompete rapacious patches.

Which type ultimately dominates? It depends on the relative force of competition within patches versus competition between patches.

Typically, more frequent local competition means that rapacious types have the edge, causing inefficient overexploitation of common resources. This tragedy of the commons occurs widely throughout nature \autocite{hardin68tragedy,leigh77how-does,frank96models,frank98foundations,rankin07the-tragedy}.

However, several conditions emphasize group against group competition, mitigating the tragedy \autocite{alexander78group,wade78a-critical,wilson80the-natural,frank22microbial}. Analysis focuses on the relative strength of forces at the different spatial and temporal scales.

\subsection{Simple model}

We can express the opposing forces of local individual competition and global group competition as
\begin{equation}\label{eq:IGtoc}
  w(x,y)=I(x,y)\,G(y).
\end{equation}
Fitness, $w$, depends on the relative success of an individual competing against neighbors in a local group, $I$, and the success of that individual's group in competition against other groups, $G$. The competitiveness of a randomly chosen focal individual is $x$, and the average competitiveness in that individual's group is $y$ \autocite{taylor96how-to-make,frank98foundations}.

For a simple example, let
\begin{equation}\label{eq:basictoc}
  w=\frac{x}{y}(1-y).
\end{equation}
Here, $I=x/y$ is the relative success of the focal individual in its group, and $G=1-y$ is proportional to the success of the group. An increase in an individual's competitiveness, $x$, gives it an advantage in local competition against neighbors \autocite{frank94kin-selection,frank95mutual}. However, greater individual competitiveness within the group degrades the group's competitiveness against other groups by raising $y$, the average amount of resources in the group that goes to competing against neighbors rather than being productive.

\subsection{Equilibrium}

Later, we will consider a variety of ways to analyze and extend this model. For now, let's use the simplest approach, highlighting the essential aspects of opposing levels and multiple scales.

At any instant in a given context, natural selection tends to increase fitness \autocite{fisher58the-genetical,price72fishers,ewens89an-interpretation,frank92fishers}. Much of the complexity in models arises from the fact that contexts change, causing what happens at different instants to change \autocite{frank15dalemberts}. If we take our context to be an evolutionary equilibrium with respect to continuously varying trait values, then an equilibrium typically occurs when fitness is at a local maximum with respect to trait values \autocite{maynard-smith82evolution}. Thus, we get a candidate for a local equilibrium by solving $\dd w/\dd x=0$ and checking for a local optimum.

In this case, $w$ depends on two variables, $x$ and $y$, so we need to use the chain rule of differentiation
\begin{equation}\label{eq:chainrule}
  \dovr{w}{x} = \povr{w}{x} + \povr{w}{y}\dovr{y}{x}.
\end{equation}
Here, $\prt w/\prt x$ measures the direct change to an individual's fitness for increasing its trait value, $x$. The term $\prt w/\prt y$ measures the change to an individual's fitness via an increase in the group average trait value, $y$. The term $r=\dd y/\dd x$ describes the change in group value, $y$, relative to the change in individual value, $x$, linking changes in $y$ back to changes in the focal individual's trait, $x$ \autocite{frank95mutual,taylor96how-to-make,frank98foundations}.

The condition for an increase in trait value is $\dd w/\dd x > 0$. Applying that condition to \Eq{IGtoc} yields
\begin{equation}\label{eq:IGtoceval}
   \prt_x\log I+r\prt_y\log IG  > 0,
\end{equation}
with equilibrium occurring when the left side equals zero. Here, $\prt_\Ga$ denotes the partial derivative with respect to $\Ga$, in which $\Ga$ can be any variable in the analysis. Note that in general $\prt_x\log f(x)=\prt_x f(x)/f(x)$, a normalized marginal change.

In this simple analysis, all trait values are equal at a local optimum. Thus, we obtain a candidate equilibrium by evaluating $\dd w/\dd x=0$ at $x=y=z^*$, with $z^*$ denoting the optimum. When we apply this method to the fitness expression in \Eq{basictoc}, we obtain
\begin{equation}\label{eq:basictocequil}
  z^*=1-r.
\end{equation}
Writing that result as a ratio of competitive tendency, $z^*$, relative to cooperative tendency, $1-z^*$, yields
\begin{equation}\label{eq:basictocratio}
  z^*:1-z^* = 1-r:r,
\end{equation}
which highlights the opposing forces acting on the trait.

\subsection{Interpretation}

Here, $r=\dd y/\dd x$ measures the slope of $y$ versus $x$. The interpretation of that slope varies with context.

In the classic tragedy of the commons analysis in biology, $r$ is the correlation in trait value between neighbors within groups \autocite{frank98foundations}. If individuals are highly correlated, then they cannot gain much against neighbors who have similar competitiveness. Thus, a lower competitiveness does better because it enhances the group component of success. Often, the more correlated neighbors are, the more they tend to cooperate to enhance the success of their group against other groups.

That way of saying things matches a typical kin selection argument \autocite{hamilton70selfish,queller92quantitative}. Genetically, the more closely related individuals are within groups, the more they tend to cooperate with each other. With respect to \Eq{basictocequil}, more related means higher $r$, less competitiveness, $z^*$, and more cooperation and better group success, $1-z^*$.

Alternatively, we can say the same thing in terms of the classic models of group selection. The correlation is
\begin{equation}\label{eq:corrvar}
  r =\frac{V_b}{V_t},
\end{equation}
in which $V_b$ is the variance in trait value between groups, and $V_t=V_w+V_b$ is the total variance as the sum of within-group and between-group variance. In other words, the correlation within groups is the same as the fraction of the total variance that is between groups \autocite{frank86hierarchical}. The ratio in \Eq{basictocratio} is
\begin{equation*}
  z^*:1-z^* = 1-r:r = V_w:V_b.
\end{equation*}

More variance between groups enhances the importance of success in group-against-group competition, favoring less competition and more cooperation. In this model, the kin selection and group selection interpretations are identical. They just use different words to say the same thing \autocite{frank86hierarchical}.

Our first example focused on horse speciation and evolution. In that case, we can think of size, $z$, as influencing both competitiveness within lineages and lineage success relative to other lineages.

However, the simple model in \Eq{basictoc} does not work for macroevolutionary pattern because it assumes the same temporal scale for competition within and between groups. For the horse example, the temporal scale of competition within species differs from the temporal scale for competition between lineages.

\subsection{Temporal scale}

We can add an adjustment to \Eq{basictoc} for different temporal scales of competition within and between groups as
\begin{equation}\label{eq:tocbasicscale}
  w=\lr{\frac{x}{y}} (1-y)^{s},
\end{equation}
in which $s$ is the frequency of competition between groups or lineages relative to competition within groups. For example, competition between species happens less frequently than competition within species, thus $s<1$. Here, $s$ enters as an exponent because fitness multiplies over time.

By the methods above, a candidate optimum for this model is
\begin{equation}\label{eq:tocscaleequil}
  z^*=\frac{1-r}{1-r+rs}.
\end{equation}
In the horse example of \Fig{williams}b, smaller size corresponds to greater competitiveness within species, whereas larger size provides an advantage in species-level selection. Thus, if we continue to think of $z^*$ as competitiveness within groups, then $1-z^*$ provides an expression for size as
\begin{equation*}
  1-z^*=\frac{rs}{1-r+rs},
\end{equation*}
which we can compare to the ratio in \Eq{basictocratio} as
\begin{equation*}
  z^*:1-z^* = 1-r:rs.
\end{equation*}
Obviously, this model oversimplifies the macroevolution of size. Nonetheless, it does capture two essential forces.

First, $r$ describes the fraction of the variance in size that occurs between lineages. If most of the variance is between lineages, then $r$ rises toward one (\Eq{corrvar}), causing size to increase because greater size corresponds to greater success of lineages. By contrast, small $r$ means that most of the variance occurs within lineages, causing the benefit of smaller size within species to dominate pattern.

Second, larger $s$ associates with more frequent lineage births and deaths, such as speciation and extinction events. More frequent competition between lineages enhances the group selection component, increasing the favored size.

Selection often happens at a variety of temporal scales and frequencies. This model can easily be extended to consider the full spectrum of selection at multiple frequencies.

\subsection{Spatial scale}

We can describe different spatial scales of competition by rewriting our basic model as
\begin{equation}\label{eq:basicspatial}
  w=\lr{\frac{x}{ay+(1-a)\bar{y}}}(1-y),
\end{equation}
a variant of an earlier model for the scale of competition \autocite{frank98foundations}.

Here, $a$ describes the scale over which individuals directly compete against others. Larger $a$ means more competition against neighbors as in our original model of \Eq{basictoc}. Smaller $a$ shifts direct competition by individuals to the global scale against randomly encountered individuals in the entire population. A random individual in that global population expresses trait value $\bar{y}$, the average over all groups.

A candidate optimum favored by selection is
\begin{equation*}
  z^*=\frac{1-ar}{1+r-ar}
\end{equation*}
or equivalently
\begin{equation*}
  z^*:1-z^* = 1-ar:r
\end{equation*}
Competition becomes more global as $a$ decreases. A lower value of $a$ reduces the amount of competition that is between similar neighbors, $ar$, causing $z^*$ to rise and individuals to become more competitive. Later, I extend this model to show how selection may happen over a variety of spatial scales.

\subsection{Symbionts, virulence, sex ratios}

The processes described in these simple models occur widely throughout life. This subsection mentions a few examples.

Symbionts living within a host may compete \autocite{eberhard80evolutionary,cosmides81cytoplasmic,hoekstra87the-evolution,hoekstra90evolution,frank96host-symbiont}. Those symbionts that reproduce faster typically transmit more rapidly to the host's progeny or to other hosts. That common scenario follows the tragedy of the commons. Direct local competition increases relative success in the host. Opposing that local benefit, greater symbiont competition may disrupt the host or reduce the symbionts' contribution to the host, degrading the symbionts' environment and reducing the success of the local group of symbionts.

Processes that reduce the correlation, $r$, between symbionts in a host favor greater symbiont competitiveness, degrading the success of the group of symbionts and perhaps also reducing the vigor of the host. Horizontal transmission of symbionts decreases $r$ \autocite{ewald94the-evolution}. In vertically transmitted symbionts, greater mutation and larger effective population size reduce $r$ \autocite{frank94kin-selection}.

The first cells in the history of life may have faced a similar challenge \autocite{szathmary87group}. Such protocells are thought to have contained a population of replicating molecules, perhaps a collection of RNA strands. Those replicators competed locally within the protocells, possibly degrading the success of the local group.

Parasite virulence follows the same natural history \autocite{levin81selection,anderson82coevolution,ewald83host-parasite,bremermann83a-game-theoretical}. The parasites compete within the host. Greater within host competition may degrade a host's vigor, reducing the group success of the parasites in that host. Virulence is the decline in host vigor caused by the parasites.

A common problem in sex ratio evolution also falls within this tragedy of the commons scenario. Suppose a few mothers lay their eggs in a restricted area. Their progeny mate within that local arena, then the mated daughters fly away to find new sites \autocite{hamilton67extraordinary}.

Let $x$ be the fraction of a mother's offspring that are sons, and $y$ be the average of $x$ in the local group. Then the relative success of a focal female's sons in local mating is $x/y$. The total success of those males increases with the number of females available for mating, which is proportional to the local fraction of progeny that are daughters, $1-y$. Then the mother's success through her sons is expressed by \Eq{basictoc}, the tragedy of the commons model \autocite{frank96host-symbiont}.

We could say that making more sons is a mother's competitive trait that degrades group success, whereas making more daughters is a cooperative trait that enhances group success \autocite{colwell81group}. Alternatively, we could say, by \Eq{basictocequil}, that the tendency to be more cooperative increases with $r$, the genetic relatedness between individuals within groups \autocite{frank86hierarchical}. The more related mothers are, the more they are favored to cooperate by reducing their investment in locally competitive sons and increasing their investment in productive daughters.

We have assumed in these scenarios that reduced local competition enhances group productivity. But in some cases the amount of local resources and potential productivity are fixed independently of the competitive and cooperative tendencies of group members. If so, then reducing local competition has no benefit through enhanced efficiency and productivity. In that case, competition tends to increase independently of $r$, the relatedness within groups \autocite{alexander74the-evolution,clark78sex-ratio,frank86hierarchical,frank86the-genetic,frank98foundations,taylor92altruism,wilson92can-altruism}.

\subsection{Dispersal}

Limited local productivity favors dispersal \autocite{hamilton77dispersal,ronce07how-does}. Consider a mother that can add a progeny to compete in her local patch or make a dispersing progeny that competes in a distant patch. Adding a competitive offspring locally degrades the success of related neighbors, discounting the competitive benefit by $r$. By contrast, a dispersing offspring will typically land in a patch with unrelated neighbors. Thus, greater relatedness among neighbors favors making more dispersers to avoid reducing the success of nearby relatives \autocite{frank86dispersal}.

A simple model illustrates how local limitation of productivity favors dispersal
\begin{equation*}
  w = \frac{1-x}{1-y+\Gd \bar{y}} + \frac{x\Gd}{1-z+\Gd \bar{y}}.
\end{equation*}
Here, $x$ is the fraction of dispersers made by a focal individual, $y$ is the average dispersal fraction in the local group, $\bar{y}$ is the average dispersal fraction in the population, and $\Gd=1-c$ is the success of dispersers that pay a cost, $c$, for dispersing. A candidate equilibrium by our usual method is
\begin{equation*}
  z^*=\frac{r-c}{r-c^2},
\end{equation*}
with $z^*=0$ for $r<c$. This model shows the powerful force favoring dispersal when local productivity is limited and neighbors are related \autocite{frank86dispersal}.

\subsection{Evolvability}

Finally, consider a simple model of time scale and evolvability, in which a lineage's success depends on a tradeoff between exploration and exploitation
\begin{equation*}
  w = \lr{1-c\Gm}\Gm^s.
\end{equation*}
Here, the trait $\Gm$ describes the tendency of a lineage to create new variants that explore alternative phenotypes. Typically, novel phenotypes degrade fitness, causing a cost, $c$, for exploration, the reduction in the trait's ability to exploit its current success. Once in a while, exploration pays off, here in proportion to the exploration rate, $\Gm$. The frequency of such payoff relative to the imposed cost against exploitation is $s$. For rare payoff, $s\ll 1$, a candidate optimum is
\begin{equation*}
  \Gm^*=\frac{s}{c}.
\end{equation*}
This result emphasizes the benefit of exploration in proportion to the frequency of payoff, $s$, relative to the cost imposed by degrading a phenotype that is currently a good one, $c$.

On the one hand, this model captures an essential aspect of how selection may shape evolvability. On the other hand, the assumptions are too simple. Typically, a trait that generates novelty, such as the mutation rate, does not itself directly cause a beneficial trait. Instead, the generative mechanism modifies other traits \autocite{karlin74towards,otto13evolution}. Thus, evolvability depends on the joint evolution of a modifying generative mechanism and a modified trait. A later section returns to this topic.

\section{Statics vs dynamics}

The previous models derived evolutionary equilibria, sometimes called evolutionarily stable strategies or ESSs \autocite{maynard-smith82evolution}. The ESS is stable against invasion in the sense that, once individuals in the population take on the ESS, any individual with a slightly different trait value will do worse. Mathematical variants consider different criteria for evolutionary stability \autocite{eshel96on-the-changing,diekmann04a-beginners,mcgill07evolutionary}. However, in most simple models, the outcome remains essentially the same.

Analyzing equilibria provides insight into statics, the balance of forces that occurs when a system is at rest \autocite{hibbeler16engineering,mankiw24principles}. Of course, the statics of a system is just a special case of the broader dynamics, so it might seem that we should always study dynamics. The problem is that to analyze dynamics, one has to make many specific assumptions about how the moving parts interact under widely varying circumstances. In any realistic evolutionary problem, we never know even approximately most of the factors involved.

We can make a lot of assumptions to analyze a mathematical model of dynamics. That teaches something. But we can be sure that the exactness of such dynamical analysis exactly describes nothing that ever really happens. Often, whether such analysis is useful or approximately true is hard to know.

People have widely varying opinions about how to use static and dynamic models to understand natural pattern \autocite{maynard-smith82evolution,samuelson83foundations,spencer05adaptive}. In my opinion, the resolution is simple as long as one keeps in mind the various costs and benefits of the different methods \autocite{frank98foundations}.

Each method provides a unique tool to understand process. In the absence of a specific problem and a particular goal, it does not make sense to say that one tool is better than another. They are different.

This article identifies the various forces of natural selection that act at different scales. Statics is typically the easiest way to identify forces \autocite{lanczos86the-variational}. However, once one has a sense of statics, it is often important to consider simple extensions to analyze dynamics. Those simple aspects of dynamics begin the task of extending one's understanding more broadly, revealing when statics fails to capture essential aspects of evolutionary process.

No matter how extensive one's analysis, there will always be another aspect of dynamics that can be added. It is a bit of a game among mathematical biologists to find yet another component of dynamics. Where should one stop? There is no right answer. But some questions help.

Does the additional complexity reveal a broadly important force that applies widely? Does it clarify some aspect of a particular problem of natural history? Does the new analysis make a prediction in a testable way?

Usually, a testable prediction takes a comparative form \autocite{frank22microbial}. For example, in our tragedy of the commons model in \Eq{basictocequil}, we derived the equilibrium trait value $z^*=1-r$. As genetic relatedness within groups, $r$, increases, individuals tend to become less competitive and more cooperative. That prediction applies widely, identifying a fundamental force that shapes traits throughout life.

Such comparisons in equilibrium models create an approach called comparative statics \autocite{silberberg74a-revision,milgrom94monotone,varian92microeconomic,samuelson83foundations}. Alternatively, we might consider how dynamics changes as some input parameter changes, comparative dynamics \autocite{judd82an-alternative,chiang92elements,kamien91dynamic}.

For example, in \Eq{tocbasicscale}, the parameter $s$ is the difference in the relative frequency of global versus local competition. The less common global competition is, smaller $s$, the bigger the ratio between the rise in competitiveness during within-group competition and the decline in competitiveness during the subsequent bout of between-group competition.

The way in which dynamics changes as $s$ changes defines a prediction of comparative dynamics. The following section illustrates this process.

\section{Dynamics of opposing forces}

This section illustrates dynamics by developing the temporal scaling model for the tragedy of the commons in \Eq{tocbasicscale}, repeated here
\begin{equation*}
    w=\lr{\frac{x}{y}} (1-y)^{s}.
\end{equation*}
In this model, we can think of $w$ as fitness over a full life cycle. Competition first occurs within groups for a relative time period of one, after which competition between groups occurs for a relative time period of $s$. We then maximize $w$ with respect to individual trait value, $x$.

When analyzing dynamics, trait values may become larger than one because of the longer periods of competition within groups and larger differences in success between competitors. To avoid fitness becoming negative through the term $1-y$, we rewrite fitness as
\begin{equation}\label{eq:tocbasicscalek}
   w=\lr{\frac{x}{y}} (\Gk-y)^s,
\end{equation}
for sufficiently large $\Gk$. The optimum for this model is simply the optimum we obtained previously in \Eq{tocscaleequil} multiplied by $\Gk$, which means that we can write
\begin{equation}\label{eq:tocscaleequilk}
 \frac{z^*}{\Gk}=\frac{1-r}{1-r+rs}.
\end{equation}
As $s$ declines, the time scale of competition within groups increases relative to the time scale of competition between groups. The greater relative time for within-group competition favors greater competitiveness, $z^*$. Put another way, for each cycle, the more generations of competition within groups relative to rounds of competition between groups, the more the short time scale of evolution within groups dominates over the long time scale of evolution between groups.

\subsection{Dynamics}

To analyze dynamics, we must make explicit assumptions. Let there be two haploid genotypes with competitiveness values $x_1$ and $x_2$ that determine fitness over a full cycle of competition within groups. The initial frequencies of the two types within a group are $q$ and $1-q$, with global frequencies $\bar{q}$ and $1-\bar{q}$. The difference in competitiveness of types is $\Ga=x_1-x_2$, and the average competitiveness within a group is $y=qx_1+(1-q)x_2$.

After competition within a group with initial frequency $q$, the new frequency in that group is
\begin{equation}\label{eq:qprime}
  q' = q + \frac{\Ga q(1-q)}{y}.
\end{equation}
If we let the probability that a group has initial frequency $q$ be $p_q$, and assume that $q$ takes on discrete values because groups have finite size, then the new global frequency is
\begin{equation}\label{eq:qbarprime}
  \bar{q}\,' = \sum_q p_q\;\!q'\lr{\frac{w_q}{\bar{w}}},
\end{equation}
with $w_q=\lr{\Gk-y}^{\,\! s}$ and $\bar{w} = \sum p_q w_q$. We weight each group's frequency after selection, $q'$, by the group's relative productivity, $w_q/\bar{w}$. That productivity determines the group's relative contribution to the global pool that reassorts to form new groups.

After each round of reassortment, the correlation between a type's competitiveness, $x$, and the average competitiveness in the group, $y$, is $r$. To reassort types between groups with that specified correlation, we use a specially designed distribution.

\subsection{Correlated binomial distribution}

When analyzing group-structured populations, we often need to consider the correlated distribution of individuals into groups. This section develops the correlated binomial, which may be broadly useful in models of groups.

Suppose the current population frequency of type 1 individuals with trait value $x_1$ is $\qbar$. Then $1-\qbar$ is the frequency of individuals with trait $x_2$. The correlated binomial splits the formation of groups into two parts. First, a fraction $\qbar$ of the new groups receives $M$ individuals of type 1, and a fraction $1-\qbar$ receives $M$ individuals of type 2. The remaining $N-M$ individuals in each group are added according to a binomial distribution.

Define $B(k;N,\qbar)$ as the binomial distribution of a random variable $k$ for a sample of size $N$ in a population with frequency $\qbar$ of type 1 individuals. Write the correlated binomial distribution as $cB(k;N,M,\qbar)$, which is
\begin{equation*}
  cB = \bar{q}*B(k-M;N-M,\qbar) + (1-\qbar)*B(k;N-M,\qbar),
\end{equation*}
in which `$*$' denotes multiplication. The probability, $p_q$, that a group starts with frequency $q=k/N$ individuals of type 1 is given by $cB$. 

The correlation of types within groups is
\begin{equation*}
  r = \lr{\frac{M}{N}}^2 + \frac{N-M}{N^2}.
\end{equation*}
The values of $M$ and $N$ are integers, so the value of $r$ must be chosen consistently with those integer constraints.

\subsection{Optimum and polymorphism}

The static analysis leading to \Eq{tocscaleequilk} predicted the optimum unbeatable trait value. Here, I show that the dynamics given by \Eq{qbarprime} support the static analysis. In particular, I tested a sample of starting values for $\qbar$ and parameters $N$, $M$, $s$, $r$, and $\Gk$. In each case, a trait value at the predicted optimum beat all other trait values and increased to a frequency of 1. The freely available software code shows the calculations \autocite{frank24natural}.

The numerical studies also showed that particular combinations of trait values, $x_1$ and $x_2$, can be maintained polymorphically at intermediate frequencies of $\qbar$. Polymorphism appears to require that $x_1$ and $x_2$ be on opposite sides of the predicted optimum, $z^*$, in \Eq{tocscaleequilk}.

\begin{figure*}[t]
\centering
\includegraphics[width=0.55\hsize]{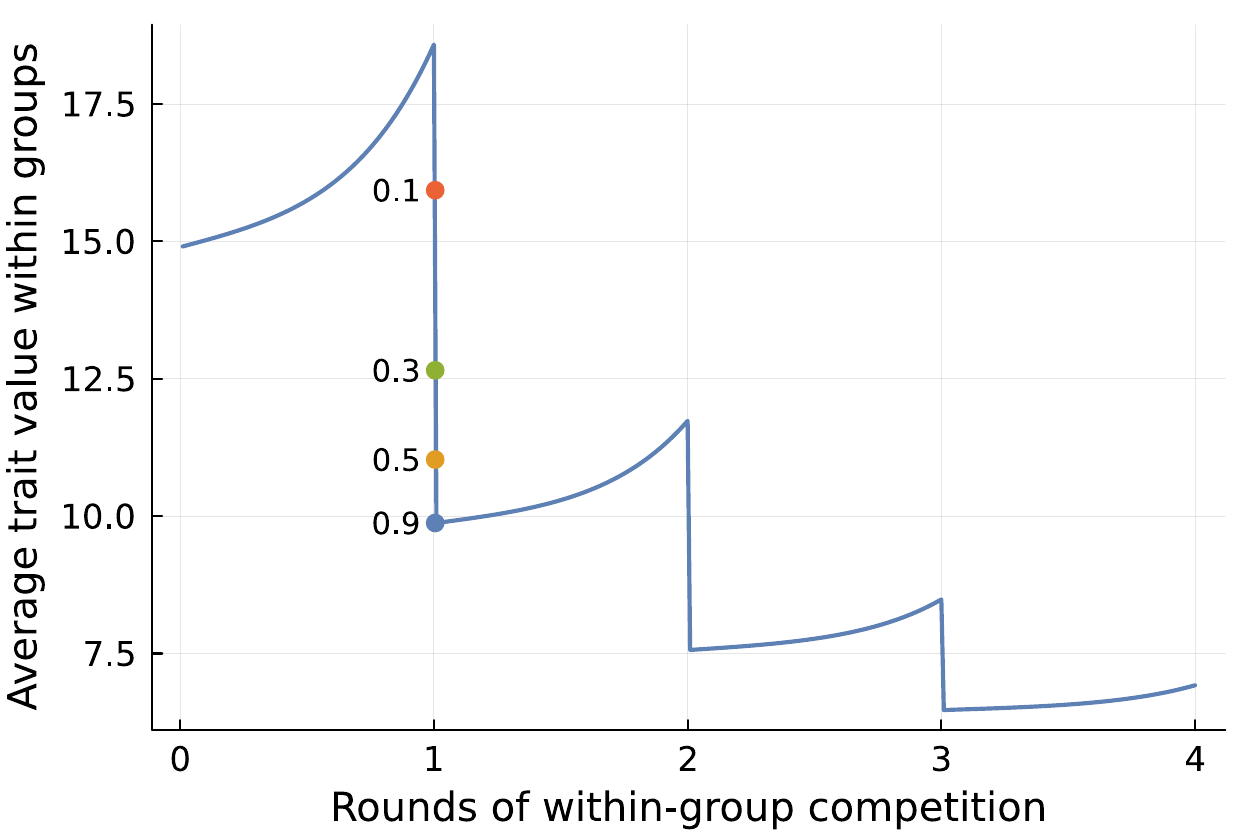}
\vskip0pt
\caption{Alternating phases of within- and between-group selection. Initially, groups are formed by the correlated binomial distribution. Then selection occurs within each group, causing a rise in the average competitive trait value. Here, the average is measured across groups without weighting by group size. That unweighted measure expresses the average competitiveness per group, considering each group equally. After each round of within-group selection, between-group competition causes a sharp drop in trait values because groups with lower competitiveness are more productive. The dynamics over a full cycle are given by \Eq{qbarprime}. Parameters for this example are $N=1000$, $\Gk=100$, initial $\qbar=0.9$, $s=0.9$, $r=0.95065$, $M=975$, which lead to an optimum of $z^*=5.453$. The competing trait values are $x_1=z^*$ and $x_2=\Gk$. The labeled circles show the amount of decline in competitiveness, for different values of $s$, during the first between-group competition period. Following the comparative dynamics prediction given in the prior section, smaller values of $s$ associate with smaller declines relative to the rise in competitiveness during the within-group competition period. The software code to produce the figure is freely available \autocite{frank24natural}.
}
\label{fig:temporalDyn}
\end{figure*}

\subsection{Phases of the cycle}

When the short time scale of competition within groups dominates, there can be significant evolution of traits within groups. Typically, competitiveness increases during the within-group phase of selection.

More intense competition within a group often reduces that group's ability to compete against other groups. Thus, less competitive groups dominate in colonizing new resource patches, causing competitiveness to drop at the start of each new cycle, after the initial colonization of groups.

Observed trait values depend on when one measures individuals. For example, many generations of viral evolution may occur within hosts between each round of transmission between hosts. Competitiveness will be relatively high after a long period of within host competition.

If viral competitiveness causes virulence and reduces host fitness, then competitiveness will be relatively low at the start of infections within hosts because, at the start of each new infection, more viruses will have come from hosts with lower virulence and less competitive viruses. The more virulent competitive viruses were more likely to harm their hosts and reduce their chance for transmission \autocite{levin94short-sighted}.

To analyze the different phases of the cycle, we can separate the dynamics during within host and between host selection. We can write the within host dynamics in frequency by following \Eq{qbarprime} and dropping the between host component, yielding
\begin{equation}\label{eq:withinPhase}
  \bar{q}\,'(t) = \sum_q p_q\;\!q'(t),
\end{equation}
in which $\bar{q}\,'(t)$ is the frequency at time $t$ averaged over all groups and not weighted by the size of the groups. As before, a group with initial frequency $q$ occurs with probability $p_q$.

To calculate $q'(t)$, the frequency in a group at time $t$ with initial frequency $q$, we use \Eq{qprime} and express the trait values in terms of the dynamics of growth as $x_i(t)=e^{m_i t}$. We set the interval for $t$ to $[0,1]$, with the value of traits used previously as $x_i=e^{m_i}$.

\Figure{temporalDyn} shows the alternating phases of within- and between-group selection. During a period of within-group selection, the more competitive type wins, causing a rise in the average value of the competitive trait. At the end of each phase of within-group competition, the groups compete for colonizing new patches. That between-group phase causes a drop in the frequency of competitive types because lower competitiveness associates with greater group success.

The fluctuating changes caused by the opposing forces can be large. In the first phase of within-group competition, the average competitiveness rises by approximately $20\%$. Then during the between-group competition phase, the average competitiveness drops by approximately $40\%$. The fluctuations in each phase decline as the population moves toward its optimum value.

The prior section mentioned a prediction of comparative dynamics. The less common global competition is, smaller $s$, the bigger the ratio between the rise in competitiveness during within-group competition and the decline in competitiveness during the subsequent bout of between-group competition. In \Fig{temporalDyn}, the circles along the initial decline during between-group competition are labeled with different values of $s$, illustrating the prediction of comparative dynamics.

\subsection{Continuous vs discrete, forces of constraint}

If continuous variation in traits arises by mutation or other processes, then eventually the optimum trait value will arise and dominate the population. Alternatively, when only a few distinct trait values occur, then the optimum trait is unlikely to occur. If at least one trait value is above the optimum and another is below the optimum, then the polymorphic combination may be maintained.

An intrinsic limit on the set of possible trait values acts as a force of constraint. That force of constraint restricts the evolutionary flow of dynamics, potentially having a strong effect on the observed pattern of trait values.

\subsection{Mutation, short-sighted evolution}

Mutation or other variance-generating mechanisms can also influence the distribution of variation within and between groups \autocite{frank94kin-selection,levin94short-sighted}. Suppose, for example, that many generations of competition happen within groups before each round of competition between groups. If the size of each group becomes sufficiently large, then highly competitive mutants may inevitably arise and come to dominate within each group.

The ultimate dominance by highly competitive types within each group greatly reduces the variance between groups. Low variance between groups is the same as a low value of the correlation within groups, $r$. Even when each group is founded by a single type and the initial correlation is $r=1$, the degradation of between-group variance by mutation and selection within groups may greatly reduce the effective correlation, favoring highly competitive types \autocite{frank10the-trade-off,frank10microbial,frank13microbial}.

Some viral life cycles may follow this scenario \autocite{levin94short-sighted}. Within each host, there may be many generations of competition, very large populations, and relatively high mutation rate. Inevitably, mutants with faster growth will come to dominate within hosts. Those faster growing variants may be more virulent, harming the host and degrading the overall potential for viruses in a host to transmit to new hosts. In essence, the increase in local competitiveness degrades the commons.

Such virulence and degradation of the commons can happen even when each host is infected by a single viral clone. In that case, all of the variance is between hosts at initial infection, and $r=1$. However, as mutation and selection happen within hosts, the difference in viral populations between hosts declines as within host competitiveness inevitably rises.

Sampling hosts when they have significant viral load leads to a measure of high average competitiveness and high virulence, even when initial infections are clonal. In other words, virulence may be high because short-sighted local evolution dominates the long time scale of group selection \autocite{levin94short-sighted}. George Williams suggested that such short-sighted evolution may dominate whenever the short time scale of local competition happens much more frequently than the long time scale of global competition between groups \autocite{williams66adaptation}.

\subsection{Explanation vs calculation}

Our simple model in \Eq{tocbasicscale} captures this essential force of relative time scale. In that model, the frequency of competition between groups relative to competition within groups declines as $s$ declines. Smaller $s$ associates with a relatively stronger force within groups, raising the optimum competitiveness, as shown in \Eq{tocscaleequil}. Similarly, greater mutation degrades $r$, also increasing the favored value of competitiveness.

The simple static models highlight essential forces rather than calculation of dynamics. The dynamical models provide more detail about how things change over time. But that extra detail imposes costs, such as the need to specify particular values for many parameters and the tendency for the overall motion of a system to hide the role of the individual forces that ultimately determine outcome. 

Schumpeter \autocite{schumpeter54history} noted that people are often inclined
\begin{quote}
to start from dynamic relations and then proceed to the working out of the static ones. \textit{But \dots\ in any field of scientific endeavor \dots} always static theory has historically preceded dynamic theory and the reasons for this seem to be as obvious as they are sound---static theory is much simpler to work out; its propositions are easier to prove; and it seems closer to (logical) essentials.
\end{quote}
Similarly, Lanczos \autocite{lanczos86the-variational} emphasized that statics ``focuses attention on the forces, not on the moving body.'' 

Some people strongly emphasize a ``shut up and calculate'' approach to science. The goal is to trace assumption to consequence, with outcomes that can be compared to what we actually observe without necessarily understanding why \autocite{dirac58the-principles,feynman67the-character,fuchs00quantum,mermin89whats}.

Others emphasize a ``what does it mean'' approach, how can we understand the fundamental forces that shape nature, with outcomes that can be compared to observations and a story to explain why. This duality between calculation and meaning became prominent during discussions about quantum mechanics \autocite{bell87speakable,bohr49discussion,einstein35can-quantum-mechanical,heisenberg58physics,bohm80wholeness,bricmont16making}. The same contrast naturally arises in every scientific discipline \autocite{callebaut12scientific,mayr82the-growth,mccloskey98the-rhetoric,rodrik15economics}.

Although ultimately we need both points of view, this article emphasizes forces and meaning. Because, as Schumpeter described, always static theory precedes dynamic theory.

\section{Fundamental forces, partial causes}

We can separate natural selection into the three forces of marginal value, correlation, and reproductive value. Each force defines a partial cause of change and dynamics \autocite{frank98foundations,frank22microbial}.

Knowing about a particular force helps to explain what happens but is not sufficient to determine outcome. For example, most explanations of physical motion depend partly on gravity. But knowing how gravity acts by itself does not determine motion. There are always other forces.

\subsection{Marginal value}

The first general force is marginal valuation, which concerns how natural selection tends to balance opposing forces. For example, being bigger may help to compete and to avoid predation. But larger size may require more resources for maintenance and trigger easier detection by predators.

The balance occurs when a small marginal increase in size causes equal gains and losses \autocite{jevons71the-theory,marshall90principles,mas-colell95microeconomic,blaug97economic}. If gains exceed losses, then selection favors an increase. If losses exceed gains, then selection favors a decrease.

Similarly, how does natural selection alter traits to balance the competitive gains against neighbors versus the cooperative efficiency of resource use? The balance typically occurs when the marginal gain for slightly better competitive success equals the marginal loss for slightly worse efficiency.

If the marginal gain in competition were greater than the marginal loss in efficiency, then selection would alter traits to enhance competition. Traits change until marginal gains and losses balance. Overall, the marginal changes in fitness define an instantaneous partial cause acting on a trait with a particular value. 

To develop an example of marginal gains and losses, recall the general form for our basic tragedy of the commons model in \Eq{IGtoc}
\begin{equation*}
  w(x,y)=I(x,y)\,G(y),
\end{equation*}
in which individual competitiveness against neighbors, $x$, depends on a trait such as resource uptake rate. The average resource uptake rate in the local group is $y$. The focal individual's share of local group success is $I(x,y)$, and the group efficiency in using resources is $G(y)$. 

Normalizing fitness to be one at the evolutionarily favored trait value, $z^*$, often helps to obtain a consistent interpretation of forces. Writing
\begin{equation*}
  w=\frac{I(x,y)}{I{\lr{z^*,z^*}}}\;\frac{G(y)}{G(z^*)}
\end{equation*}
yields a normalized fitness of one when evaluated at the fixed point, $x=y=z^*$.
  
We obtain the trait value that maximizes fitness by following the steps in \Eq{chainrule} and setting the derivative equal to zero, yielding
\begin{equation*}
  \dovr{w}{x}=-\Cm+r\Bm=0,
\end{equation*}
with
\begin{equation*}
  -\Cm=\frac{I_x}{I}\mskip50mu \Bm=\frac{I_y}{I}+\frac{G_y}{G},
\end{equation*}
in which a subscript means a partial derivative with respect to that variable. All functions are evaluated at the fixed point, $x=y=z^*$. The marginal costs and benefits equalize at a candidate for a maximum, yielding
\begin{equation*}
  C_m=rB_m.
\end{equation*}

In our model of temporal scale in \Eq{tocbasicscalek}, $I=x/y$ and $G=(\Gk-y)^s$. Here, $x$ and $y$ describe competitiveness. If we analyze $-\dd w/\dd x$, then the marginal values describe the tendency for efficiency, group benefit, and cooperation. Near the optimum, $z^*$, the marginal costs and benefits of cooperation are
\begin{equation}\label{eq:CmBm}
  \Cm=\frac{1}{z^*}\mskip50mu \Bm=\frac{1}{z^*} + \frac{s}{\Gk-z^*},
\end{equation}
which leads to the optimum in \Eq{tocscaleequilk}.

For any trait that can take on continuous values, an optimum typically balances marginal costs and benefits.

\subsection{Correlation, group selection, and kin selection}

The second general force is correlation between interacting individuals or other biological units. This subsection illustrates how different biological problems lead to similar looking mathematical expressions for correlation. Once one recognizes the few ways in which patterns of association and variance influence natural selection, one can distinguish the small number of seemingly similar but actually distinct causal processes that commonly arise \autocite{frank98foundations}.

Mathematical analyses of natural selection often lead to expressions that include covariance terms \autocite{robertson66a-mathematical,price70selection,price72extension}. In different scenarios, those covariances may arise from phenotypic correlations between group members, variances between groups, or genetic relatedness between interacting individuals. The covariance's promiscuity leads to confusion about the ways in which various causal processes arise in natural selection \autocite{frank97the-price,wolf99interacting,frank12naturalb}.

Let's start with the basic tragedy of the commons model
\begin{equation*}
  w = \frac{x}{y}(1-y).
\end{equation*}
Consider a population with many independent groups. We randomly choose a focal group, with average trait value $y$. In that group, we choose a random individual with trait value $x$. In this model, nothing distinguishes individuals or groups except their trait values. Everyone is a member of the same class.

Recall that we derived the optimum trait value to be $z^*=1-r$, in which $r$ is the correlation between a randomly chosen individual's trait value and the average trait value of that individual's group. This correlation determines the outcome. How should we interpret the meaning of this correlation in terms of biological processes?

We start by looking at how $r$ arose in the derivation of the optimum value. In that analysis, our focal individual's fitness changes with a change in its trait value, arising from the derivative $\dd w/\dd x$ by the chain rule given in \Eq{chainrule}. We then defined $r$ to be the component slope $\dd y/\dd x$, the change in the average trait of the focal individual's group, $y$, as its own trait, $x$, changes. We called $r$ the correlation of an individual to its local group, in which the local group average includes the focal individual's contribution to that group average.

More comprehensive analysis shows that the actual slope that we need is the regression coefficient of $y$ on $x$ \autocite{queller92quantitative,taylor96how-to-make,frank97multivariate}, which is
\begin{equation*}
  \byx = \cov(x,y)/\var(x).
\end{equation*}
In the limit of small changes in $x$, this slope becomes the derivative of $y$ with respect to $x$. When searching for a local optimum, we can use the small change expression of the derivative. However, to understand the meaning of $r$, it is useful to analyze the more general regression expression.

For a group size of $N$, the group average is $y=\sum x_i/N$. Using that definition of $y$ in the regression expression leads to
\begin{equation*}
  \byx=\frac{1}{N}+\frac{(N-1)\Gr}{N}=r,
\end{equation*}
in which $\Gr$ is the correlation between randomly chosen pairs of different individuals within a group, and $r$ is the correlation between randomly chosen pairs of individuals sampled from a group with replacement. In other words, $r$ includes the probability $1/N$ that two individuals sampled with replacement will be identical.

It is easy to think of the basic tragedy model as describing groups of individuals sampled from the same species. However, the model is more general. Different groups may represent different species. Or each group may be comprised of individuals from different species, for example, a mixture of different bacteria and yeast cells competing for a common resource \autocite{frank94genetics}. 

From that wider perspective, four distinct biological processes may influence the regression coefficient, $\byx$. We must clearly separate those distinct processes in order to understand how natural selection shapes traits. Much of the challenge and controversy arises from the failure to separate those distinct processes, partly hidden by the seemingly similar regression and correlation coefficients that appear in each case.

\subsubsection{Correlated traits}

Suppose groups include a mixture of microbial species. Each individual in the group has a competitive tendency, $x$, to suck up local resources. As the local average, $y$, rises, the greater allocation of individual resources toward competitiveness reduces the overall efficiency of the group in reproducing and transforming resources into biomass.

The tragedy model applies. The more closely correlated individuals are in their competitiveness trait, the more natural selection will reduce local competitiveness and enhance group efficiency. It is easiest to imagine how this would work when the individual trait values are influenced by genotype. Then the similarity defined by the regression $\byx$ depends on the genetic associations between the different species within groups \autocite{frank94genetics}.

However, the causal process ultimately depends on phenotypic similarity and does not require genetic similarity. For example, consider two individuals that form a group. Focus on one. If its partner expresses a correlated level of competitiveness, then a rise in the focal individual's competitiveness leads to relatively little gain because the partner will typically also have a higher competitiveness.

Against that small gain from enhanced competitiveness, the cost rises significantly because both individuals have increased their allocation to competitiveness, deteriorating the local efficiency of the group. Although easiest to imagine when traits have a genetic basis, that genetic aspect is not necessary. Only the phenotypic correlation is required. In this case, the process follows Aumann's correlated equilibrium ideas from game theory \autocite{aumann74subjectivity,skyrms96evolution,frank97the-price,frank98foundations,moore97interacting,wolf99interacting}.

Of course there has to be some heritable tendency for the expression of traits. Otherwise, success does not translate into evolutionary change. However, the heritable transmissibility of traits may differ from the factors that cause phenotypic associations, as shown in the discussion below on inclusive fitness \autocite{frank97the-price,moore97interacting,wolf99interacting}.

\subsubsection{Group selection}

The primary causal role of regression or correlation does not depend on group structure. What matters is the scale that sets the association of interacting individuals in relation to the scale of competition for resources \autocite{frank98foundations}.

Group structure may impose the same scales on competition and resource efficiency, making it easy to understand process. Thus, group selection provides a convenient label when clear boundaries occur but obscures the underlying causal processes in other situations \autocite{hamilton75innate}.

We mentioned earlier that in the simple tragedy model, $r=V_b/V_t$, the fraction of the total variance that occurs between groups. Although correct for certain situations, this again obscures causal process. The correct interpretation of $r$ is the regression $\byx$. However, for that value, we must ultimately understand whether a particular situation arises from purely phenotypic associations or from an underlying genetic basis.

\subsubsection{Neighbor-modulated kin selection}

Consider again the scenario in which groups comprise a mixture of microbial species. Suppose any phenotypic correlation has a genetic basis. The particular genes and the mechanisms that link genotype to phenotype may differ between species. If the total genetic variance is the same in each species, then the result of the tragedy model depends on $r$, the correlation between individual and group phenotype.

In this case, we are following the direct fitness of a focal individual. Neighbor phenotypes modulate the fitness of our focal individual through their effects on competition and efficiency.

When individuals are members of different species, we would not typically call this process kin selection. Yet the identical scenario with members of the same species would commonly be called kin selection or neighbor-modulated kin selection \autocite{taylor07direct,gardner11the-genetical,mcglothlin10interacting,queller11expanded}. Even for the scenario within species, the genetic and phenotypic similarities do not have to arise by interactions between kin in the usual sense of kinship by recent common ancestry.

Instead, all that matters is phenotypic similarity, whether genetic or not and, if genetic, whether truly based on kinship by recent common ancestry.

Of course, evolutionary change over time does depend on heritability, typically a genetic aspect of traits. The next section analyzes the heritable contribution of different individuals to the future population.

\subsubsection{Different classes}

To analyze heritable contributions and the transmission of traits, we must first consider the different roles that individuals may play in various scenarios. Often, it makes sense to think of different individuals with different roles as members of different classes. Then we must consider how trait values are transmitted by each class, the different pathways of heritability \autocite{taylor96how-to-make,frank98foundations}.

In our previous models, each individual was an actor expressing the same type of trait. And each individual was also a recipient, influenced by the expression of that trait in its neighbors. We had one class of individuals, each individual being both an actor and a recipient.

In other situations, a recipient may not express any trait, with its fitness influenced only by the actors' traits. For example, parents express various traits that help their offspring, whereas offspring may be purely recipients of parental benefits.

Consider an extension to our basic tragedy model, starting with the simplified linear form in \Eq{basictoc}. Here, we have two classes of individuals. Class 1 is like our prior models, in which all individuals are actors that express a trait and are recipients influenced by that expression, yielding the class fitness as
\begin{equation}\label{eq:W1}
 W_1=\lr{\frac{x}{y}}\lr{\frac{1-y}{{\;\,1-z^*}}}.
\end{equation}
In class-based models, we normalize the class fitness to be one at equilibrium so that we can compare the fitness of different classes on the same scale. Here, the extra denominator term $1-z^*$ provides that normalization.

For class 2, individuals receive the effects of class 1's trait expression but do not themselves express traits. We may think of class 1 as productive individuals and both classes as recipients of the products. In this case, class 2's fitness is
\begin{equation}\label{eq:W2}
 W_2=\frac{1-y}{{\;\,1-z^*}}.
\end{equation}

In a class-based model, we write an overall expression for fitness as
\begin{equation}\label{eq:classW}
 W = \sum_i c_iW_i,
\end{equation}
in which $W_i$ is the fitness of class i individuals, and $c_i$ weights each class by its reproductive value, the relative contribution to the future population. The next section discusses the reproductive value weightings.

For a model with classes, full analysis requires a bit of complicated notation and discussion \autocite{frank98foundations}. Here, we provide a simplified summary that highlights the most important concepts.

Previously, we evaluated the change in a trait by evaluating $\dd w/ \dd x$, the change in the fitness of a single class, $w$, with respect to the trait value, $x$. With different classes, we need to track the heritability of trait values to the future population by following the different pathways of transmission for the trait through the fitnesses of the different classes of individuals, $W_i$.

Label the transmissible value of traits $g'$, in which $g$ is for the genetic and other transmissible determinants of traits, and the prime denotes the value of those factors in the future population. Now we need the change in fitness with respect to the change in the transmissible value of the trait.

\begin{equation}\label{eq:classdW}
 \dovr{W}{g'} = \sum_i c_i\dovr{W}{x}R_i.
\end{equation}
The $R_i$ quantify the transmissible values through the different classes. There are two distinct ways for tracking those pathways of transmission.

\subsubsection{Direct and inclusive fitness}

First, in the direct fitness approach that we have been using, an $R$ coefficient measures the slope of an actor's phenotype relative to the recipients' transmissible genetic value. In this case, the pathway is a change in actor's phenotype that causes a change in recipients' fitness that causes a change in the recipients' transmissible factors. Those factors transmitted by the recipients are associated with the actor's phenotype. This pathway is direct because we measure how each recipient class directly passes transmissible factors to the future \autocite{frank97the-price,frank98foundations}.

Second, in inclusive fitness, an $R$ coefficient measures the slope of a recipient's genetic value relative to the actor's genetic value. In this case the pathway reverses, becoming a change in actor's phenotype causes a change in recipients' fitness which causes a change in the amount of actor's genetic value that is transmitted by the recipients. This pathway is inclusive to the actor by measuring all transmission with respect to the actor's genetic value \autocite{frank97the-price,frank98foundations}.

In both cases, the $R_i$ provide measures of heritability of the trait associated with each class. Overall, the direct fitness method is more general because it can handle multiple actors, and it avoids inclusive fitness's special assumptions needed to reverse the direction of causality \autocite{frank97multivariate}. For some situations, inclusive fitness may be more appealing intuitively because it assigns all consequences of transmission to the actor that causes changes in fitness.

Using the specific models for $W_1$ and $W_2$ in \Eqq{W1} and \ref{eq:W2}, a candidate for the optimal trait value occurs when
\begin{equation*}
  c_1R_1\lr{r\Bm-\Cm} + \frac{c_2 R_2 r}{1-z^*} = 0,
\end{equation*}
in which $\Bm$ and $\Cm$ are given in \Eq{CmBm}. Solving yields
\begin{equation}\label{eq:rvfixed}
  z^*=\frac{1-r}{1+\Gf r},
\end{equation}
in which $\Gf=c_2R_2/c_1R_1$. As the reproductive value and transmissibility of the second class increase, causing $\Gf$ to increase, the favored competitiveness of the actors, $z^*$, declines because less of the transmission happens through the competitive interactions in class 1.

This analysis shows a significant problem with group selection interpretations. In this two-class tragedy model, we have to track several causal factors that have no easy interpretation in terms of groups. Instead, we simply followed the pathways of causality and the weightings for various factors such as reproductive value. That simple interpretation arises naturally from identifying the various forces and tracing their consequences.

\subsection{Reproductive value}

Reproductive value weightings describe the contribution to the future population by different classes, the third general force in our analysis \autocite{fisher58the-genetical,taylor90allele-frequency,charlesworth94evolution}. This subsection introduces the basic ideas and some simple methods of analysis. The following examples show the essential role of reproductive value in evaluating temporal and spatial scales.

In \Eq{classW}, we showed how to combine different classes of individuals into an overall expression for fitness. A class that contributes relatively little to the future population will have relatively little influence on how traits evolve. The example that led to \Eq{rvfixed} showed how the relative reproductive values of different classes determine which class dominates the evolution of trait values.

\subsubsection{Traits influence reproductive value}

In the prior example, the reproductive value weightings, $c_i$, are independent of the trait value. More realistic and challenging problems arise when the reproductive values depend on the trait of interest. To study that situation, we analyze the consequences of trait values for each class and then combine the results into an overall effect.

The fitness consequence for each class depends on that class's contribution to the future population. The contribution has three aspects. 

First, the proportion of individuals in class $j$ influences the contribution of that class. We write $u_j$ for the frequency of class $j$.

Second, when class $j$ individuals contribute to class $i$, the value of that contribution must be weighted by the reproductive value of individuals in class $i$, written as $v_i$.

For example, if class $i$ represents dispersing individuals, then we must weight the contribution to class $i$ by the expected relative contribution of a disperser to the future population.

Third, the relative contribution of class $j$ to class $i$ is $w_{ij}$. For example, we may be interested in $w_{ij}(x,y)$, expressing the effect of an individual's trait, $x$, and the group average trait, $y$, on the contribution from $j$ to $i$.

The overall fitness valuation for the contribution of class $j$ to class $i$ is $v_iw_{ij}u_j$. Summing all transitions yields
\begin{equation}\label{eq:wRV}
  W = \sum_{ij} v_iw_{ij}u_j=\bmr{vAu}.
\end{equation}
Here, $\bmr{v}$ is the row vector of reproductive values per individual for each class, $\bmr{u}$ is the column vector of class frequencies, and $\bmr{A}$ is the matrix of $w_{ij}$ fitness values \autocite{taylor96how-to-make,frank98foundations}.

\subsubsection{Dispersal vs survival}

Consider a tradeoff between the production of dispersing progeny and the future survival in the current habitat \autocite{frank98foundations}, following section 5.6 of Frank \autocite{frank22microbial}.

This example has two classes. Dispersers that successfully colonize a new patch form class 1. The new colonizers and their nondispersing descendants form class 2.  Let the fitness components be 

\begin{equation}\label{eq:dispSurvA}
 \bmr{A}=
\begin{bmatrix}
 		0&\Gb(x)/D\\
 		1-d&1-\Gd-y
\end{bmatrix}.
\end{equation}
Entries in row $i$ and column $j$ denote $w_{ij}$, the contribution of class $j$ individuals to class $i$. Thus, $w_{11}$ is zero because newly arrived colonizers of class 1 do not make dispersers but instead survive locally at rate $w_{21}=1-d$ to form the surviving lineage of colonizers as class 2.

The component $w_{12}=\Gb(x)/D$ describes the contribution of the local lineage to dispersers that successfully colonize a new patch. The local lineage's investment in making dispersers is $x$, and $\Gb(x)$ is the functional relation between dispersal investment and dispersal success. Dispersal success is normalized by the density-dependent factor, $D$, in which greater density-dependent limitation reduces dispersal success.

The component $w_{22}=1-\Gd-y$ describes the survival of the colonizing lineage within its patch. The intrinsic loss rate is $\Gd$, which combines destruction of the patch, loss of the colonizers from a continuing resource patch, or death of a host when the colonizers are parasites. 

The intrinsic loss rate is increased by $y$, which is the patch average of the trait value $x$ that determines the number of successful dispersers. As successful dispersal rises, the local survival rate decreases, similar to a tragedy of the commons model.

When evaluating total fitness, $W$, from \Eq{wRV}, we need the individual reproductive values, $\bmr{v}$, for the classes when evaluated at demographic equilibrium, $x=y=z^*$, derived in Frank \autocite{frank98foundations} as
\begin{equation}\label{eq:vdisperse}
 \bmr{v}\propto\begin{bmatrix}1-d&\Gl\end{bmatrix},
\end{equation}
in which ``$\propto$'' means \textit{proportional to}. The reproductive value of new colonizers is discounted by $1-d$, which is the probability of surviving the initial delay after colonization and before producing dispersers. The reproductive value of residents is augmented by $\Gl$, the population reproduction rate because residents have average reproductive success $\Gl$ during the period when new colonizers do not reproduce.

The value of $\Gl$ is the dominant eigenvalue of the fitness matrix $\bmr{A}$ evaluated at $x=y=z^*$. Later, we will deal with the fact that $\Gl$ depends on the other model parameters. For now, we consider the population's average reproductive rate as a separable partial cause.

The class frequencies at demographic equilibrium are proportional to
\begin{equation}\label{eq:udisperse}
 \bmr{u}\propto\begin{bmatrix}\Gb(z^*)/D\\ \Gl\end{bmatrix}.
\end{equation}
To obtain the trait values that maximize the total fitness in \Eq{wRV}, we evaluate $\dd W/\dd x=0$ at $x=y=z^*$, which includes
\begin{equation*}
 \dovr{\bmr{A}}{x}=
\begin{bmatrix}
 		0&\Gb'(z^*)/D\\
 		0&-r
\end{bmatrix}
\end{equation*}
and the vectors $\bmr{v}$ and $\bmr{u}$ at demographic equilibrium, leading to a solution that must satisfy $v_1\Gb'(z^*)/D=v_2r$ at which the marginal gains and losses for dispersers are equal, yielding
\begin{equation*}
 \Gb'(z^*)=\frac{r\Gl D}{1-d}.
\end{equation*}
If we assume that dispersal success is $\Gb(z)=z^s$, with $s<1$, then dispersal success rises at a diminishing rate with investment in dispersal, yielding the solution
\begin{equation}\label{eq:dispSurv1}
 z^*=\lrb{\frac{s(1-d)}{r\Gl D}}^{1/1-s}.
\end{equation}
The various terms interact to determine the favored dispersal rate, $z^*$. We can get a sense of partial causation by considering how $z^*$ changes in response to partial changes in the terms. In particular, a rise in $d$ lowers the initial survival of colonizers within a patch, decreasing investment in dispersal. Similarly, a rise in $\Gl$ raises the growth of patch residents, lowering the relative value of colonizers and also decreasing investment in dispersal.

A decrease in density-dependent limitation, $D$, increases the opportunity for dispersers to find new patches, raising dispersal. Smaller values of $s$ cause more rapid saturation of dispersal success, lowering dispersal investment.

This model also expresses the tragedy of the commons. Reduced similarity, $r$, favors more dispersal, which decreases local survival and the long-term quality of the local patch. In other words, dispersal is a competitive trait that degrades the local commons by more rapidly extracting local resources to develop dispersal-enhancing traits.

In this example, different classes have the same trait heritabilities. When those heritabilities differ by class, we must weight each class's success by its relative heritability for the trait, as in the $R_i$ coefficients of \Eq{classdW}.

These conclusions provide a rough qualitative sense of how various forces shape dispersal. In each case, we emphasized how a change in some factor leads to a partial pathway of causation favoring either an increase or a decrease in dispersal.

\subsubsection{Scale}

The dispersal example illustrates natural selection's scaling of time and space by reproductive value. Temporally, in \Eq{vdisperse}, the ratio of individual reproductive values, $\Gl/(1-d)$, describes the benefit gained by older reproductive individuals relative to younger colonizing individuals that must survive one time step before reproducing.

The value of $\Gl$ describes population growth, with $\Gl=1$ at steady population size. Greater population growth more strongly emphasizes the current reproduction of nondispersers over the future reproduction of dispersers.

Current reproduction benefits from population growth because population size after $t$ steps into the future is proportional to $\Gl^t$. Thus, adding a new individual to the population at time $t$ provides a contribution in proportion to $\Gl^{-t}$.

A single addition to a bigger population is a smaller fractional contribution to the future composition of that population. In this case, the reproductive value ratio $\Gl/(1-d)$ provides a simple measure for the scaling of time.

Spatially, in \Eq{udisperse}, the number of successful dispersers declines with increasing density dependence, $D$, degrading the benefit of spatial movement. In effect, natural selection increasingly discounts reproductive value by distance of movement, the role of spatial scale. Similarly, larger $\Gl$ and more rapid population growth increase the relative number of resident reproductives, valuing more strongly the local versus distant scale for future contribution to the population.

These interpretations of scale follow from the population reproductive rate, $\Gl$. We mentioned earlier that $\Gl$ depends on the other model parameters. That fact complicates the interpretation of the model with respect to its individual parameters.\

To clarify things in analysis, one typically makes the reasonable assumption that $\Gl=1$, meaning that population size is in steady state when the trait of interest is at its optimum value, $x=y=z^*$. This assumption causes the density dependence term to become
\begin{equation*}
  D=\frac{(1-d)\Gb(z^*)}{\Gd+z^*}.
\end{equation*}
This expression simplifies the reproductive value expressions and leads to an optimum trait value of
\begin{equation*}
  z^*=\frac{s\Gd}{r-s}
\end{equation*}
for $r(1-\Gd)>s$.

However, by simplifying in this way, we lose insight into how natural selection discounts reproduction in relation to scale by the factor $\Gl$, the population rate of reproduction. 

As often happens, we face the common tradeoff between isolating partial causes in a clear and intuitive explanatory way versus calculating the relations between particular model parameters and specific outcomes. Ultimately, explanation and calculation are important in different ways, so it helps to see things from both points of view.

\section{Temporal scaling}

In the previous section, current reproduction had different consequences from future reproduction. A growing population favors immediate reproduction over future reproduction. A shrinking population favors later reproduction over current reproduction. Overall, the intensity of natural selection acting on a trait scales with time.

This section presents a more general tragedy of the commons model that highlights temporal scaling \autocite{frank10demography}. Here, we follow Section 5.7 of Frank \autocite{frank22microbial}.

Suppose some individuals colonize a resource patch and grow for many generations. They also send dispersers to colonize other patches. Those dispersers can be thought of as the reproduction or fecundity of the group. Total reproduction over the colony life cycle depends on how long the colony survives.

We must consider, at each temporal stage in the colony life cycle, how traits influence an individual's relative share of the group's current and future genetic transmission. We multiply that reproductive share by the total productivity of the group.

\subsection{Cycle fitness}

In this case, we write a single expression that combines the fecundity and survival components of fitness over the full life cycle. A colony grows through $j=0,1,\ldots$ temporal stages. The fitness of a focal individual in the $j$th stage is
\begin{equation}\label{eq:cycleFitTr}
  w_j=I\lr{x_j,y_j}\sum_{k=j}^{\infty}\Gl^{-k}\,G\lr{\bmr{y}_k}.
\end{equation}
The first term, $I$, describes an individual's share of the colony's long-term success. In a tragedy model (\Eq{IGtoc}), $I$ increases with an individual's competitive trait expression, $x_j$. For example, $I=x_j/y_j$ expresses the relative competitive success of an individual with trait $x_j$ when competing in a group with average competitive trait value, $y_j$.

The second term describes the reproductive value for the colony in the $j$th stage. That value is the sum of the colony success, $G$, in the current stage, $j$, and in all future stages. Colony reproductive value at stage $k$ depends on $\bmr{y}_k=y_0,y_1,\ldots,y_k$, the average trait value at each stage up to and including the current stage. 

The colony success for each stage is multiplied by the discount for the amount the population size has grown, $\Gl^{-k}$, since colony inception at stage $j=0$. We discount future reproduction by the expansion of the population size because a single progeny represents a declining share in an expanding population. In the following analysis, $\Gl=1$, so that we can highlight other processes that influence temporal scaling and reproductive value.

The group success in stage $k$ can be divided into survival and fecundity components of reproductive value,
\begin{equation*}
  G\lr{\bmr{y}_k}=S\lr{\bmr{y}_k}F\lr{\bmr{y}_k}.
\end{equation*}
The survival to stage $k$ is $S\lr{\bmr{y}_k}$, and the fecundity is $F\lr{\bmr{y}_k}$. We find the trait vector, $\bmr{z}^*$, that maximizes fitness by simultaneously evaluating $\dd w_j/\dd x_j=0$ for all $j$ when evaluated at $\bmr{x}=\bmr{y}=\bmr{z}^*$.

\subsection{Colony tragedy of the commons}

Suppose the colony grows without producing dispersers from generations $k=0,1,\dots,g-1$. Then surviving colonies remain at constant size and produce migrants in proportion to their fecundity in each of the following generations.

With those assumptions, the components of individual success, group survival, and group fecundity are, respectively
\begin{align*}
  I\lr{x_j,y_j}&=\frac{x_j}{y_j}\\
  S\lr{\bmr{y}_k}&= S\lr{\bmr{z}_k^*}\lrb{\frac{1-y_j}{1-z_j^*}}^{\Gth(g-1-j)}\\[6pt]
  F\lr{\bmr{y}_k}&= F\lr{\bmr{z}_k^*}\lrb{\frac{1-y_j}{1-z_j^*}}.
\end{align*}
Individual success, $I$, follows the standard tragedy model. An individual's share of group success in the $j$th generation is the ratio of its competitive trait, $x_j$, relative to the group average, $y_j$.

Survival to generation $k$ depends on the survival in each of the preceding generations. Thus, any cooperative enhancement of survival in a particular generation carries a benefit forward to all future generations. In this model, deviations in group trait values only influence survival during the juvenile generations, $j<g-1$.

In each juvenile generation, $j$, the survival consequence of a deviation in group trait value, $y_j$, is $[(1-y_j)/(1-z_j^*)]^\Gth$. That value multiplies for each of the $g-1-j$ juvenile generations over which it acts. Any consequence to total survival over the juvenile period also affects cumulative survival to future reproductive generations. The value of $S\lr{\bmr{z}_k^*}$ is the baseline survival rate to generation $k$ in a group without deviant trait values.

The fecundity consequence for a deviation in group trait value is $(1-y_j)/(1-z_j^*)$. The value of $F\lr{\bmr{z}_k^*}$ is the baseline fecundity in generation $k$ in a group without deviant trait values.

Solving $\dd w_j/\dd g_j=0$ for all $j$ when evaluated at $\bmr{x}=\bmr{y}=\bmr{z}^*$ yields $z_j^*$, the favored trait value in each generation $j$. When expressed as the competitive to cooperative tendency, $z_j^*:1-z_j^*$, we obtain
\begin{equation*}
  1-r:r\lr{1+\Gg_j},
\end{equation*}
with the enhanced demographic component for the cooperative tendency caused by the trait's contribution to colony survival as
\begin{equation*}
  \Gg_j=
    \begin{cases}
      \Gth(g-j-1)&j<g-1\\
      0&j\ge g-1.
    \end{cases}
\end{equation*}
This model illustrates the increased selective force on cooperative and competitive traits during the early stages of colony growth when $j$ is small. Put another way, selective force scales with time.

\section{Spatial scaling}

An earlier model in \Eq{basicspatial} introduced two scales of competition, local and global. This section extends the analysis to include varying spatial scales for both competition and cooperation.

Let fitness be
\begin{equation*}
  w = \frac{x}{\sum a_iy_i}\lr{1-\sum b_iy_i},
\end{equation*}
in which all sums run over $i=0,1,\dots,n$, with $0$ corresponding to local interaction and each increasing step associating with greater distance away from the focal patch. The $a_i$ and $b_i$ coefficients describe the probability of interaction at each distance for, respectively, competitive and cooperative components.

Our standard methods lead to a simple extension of earlier results
\begin{equation*}
  z^* = \frac{1-\sum a_ir_i}{1-\sum \lr{a_i-b_i}r_i},
\end{equation*}
in which $r_i$ is the correlation in trait value between an individual in a focal patch and the average trait value at $i$ steps in distance from that patch. We can write this expression as a ratio $z^*:1-z^*$ of competitive to cooperative tendency in trait values
\begin{equation*}
  1-\sum a_ir_i : \sum b_ir_i.
\end{equation*}
This result illustrates how the variety of spatial scales influences trait values.

Once again, we see that group selection would be a difficult concept to apply because there are no set group boundaries for competition and cooperation, which happen over different spatial scales.

A different type of spatial scaling arises when habitats vary in quality. If individuals can adjust their trait expression to match each habitat, then we can treat each habitat type as posing an independent evolutionary challenge. For example, if the quality of patch  $i$ is $\Gk_i$, and the average trait value in that patch is $y_i$, then
\begin{equation*}
  w_i = \frac{x}{y_i}\lr{\Gk_i-y_i}
\end{equation*}
and, as in \Eq{tocbasicscalek}, we have
\begin{equation*}
  z_i^*=\Gk_i\lr{1-r_i}.
\end{equation*}
Alternatively, if individuals cannot adjust their traits in response to the local conditions of patch quality, $\Gk_i$, and correlation with neighbors, $r_i$, then fitness is averaged over patches of different quality and correlation
\begin{equation*}
  w=\sum \Gg_iw_i
\end{equation*}
for different patch types $i$ that occur with frequency $\Gg_i$. Then
\begin{align*}
  z^* = \sum \Gg_iz_i^* &=\sum \Gg_i\Gk_i\lr{1-r_i}\\[5pt]
  						&=\bar{\Gk}\lr{1-\bar{r}}-\cov\lr{r,\Gk},
\end{align*}
in which overbars denote averages over the patch frequencies, and the final term is the covariance between the correlation of trait values within a patch, $r$, and patch quality, $\Gk$.

\section{Multiple traits in communities}

Different species often influence each other's fitness. For a pair of species, individuals interact over a particular spatial scale. Then, those interacting individuals, or their descendants, may move together or separately over various spatial scales. The scales of interaction and cotransmission influence how traits evolve \autocite{frank94genetics,foster06a-general,leigh-jr10the-evolution,sachs04the-evolution}.

\subsection{Two-species tragedy of the commons}

As an example, consider the fitnesses for each of two species in an extended tragedy of the commons model
\begin{equation*}
  w_i = \frac{x_i}{y_i}\lr{1-y_1}\lr{1-y_2}\qquad i=1,2.
\end{equation*}
Species $i$ competes directly only with other members of the same species, with locally competitive trait $x_i$ relative to the average of that competitive trait, $y_i$, of the same species.

The average competitive trait of species $i$ degrades the local environment by $1-y_i$. Each species is influenced by the degradation caused by both species, which combine multiplicatively.

We are interested in how the interaction between species may influence competitiveness and the degradation of the commons. The outcome depends on the cotransmission of traits between the two species. For a pair of traits that work particularly well together, to what extent do those trait variants tend to stay together in subsequent interactions?

To isolate the forces acting in that situation, we must evaluate how selection influences the direction of change in trait values in each species. In particular, we evaluate $\dd w_i/\dd x_i$, the change in the fitness of an individual in the $i$th species relative to a change in its trait value.

When $\dd w_i/\dd x_i<0$, selection favors a reduction in the competitive trait, which is equivalent to an increase in the tendency for cooperation and group success. To reduce the notational complexity and highlight how particular forces act as partial causes, we evaluate the derivative at $x_i=y_i=z$, which means that all individuals in both species have traits near to the same value, $z$.

If we also assume that the trait variance in each species is the same, then the condition that favors a decrease in competitive tendency is
\begin{equation*}
    z > \frac{1-r}{1+\Gr}.
\end{equation*}
We can write the same result in terms of cooperative tendency, $1-z$, with the condition that favors an increase in cooperation as
\begin{equation*}
  1-z < \frac{r+\Gr}{1+\Gr},
\end{equation*}
in which $r$ is the correlation in trait value between individuals of the same species within the spatial scale over which competition and resource degradation occur, and $\Gr$ is the correlation in trait values between interacting species with respect to that same spatial domain.

\subsection{Scale of selection relative to cotransmission}

The novel factor influencing cooperation arises from the correlation between species, $\Gr$. Two opposing forces act on that correlation \autocite{frank94genetics,frank95the-origin,frank97models}.

First, selection tends to increase the between-species correlation. Pairs of trait values that work well together increase because of their greater success. Pairs that work poorly together decrease. In this case, a group with lower trait values for competitiveness produces more descendants than a group with higher trait values. That synergism between species causes a positive correlation in trait values.

Second, when individuals from the two species disperse independently, the recombining of individuals to form new groups breaks up correlated combinations of trait values. Alternatively, cotransmission of individuals and traits to form new interacting groups maintains the correlations in trait values created by selection.

The balance between selection and recombination determines the correlation between species, $\Gr$. Many particular assumptions influence that balance. The details are beyond our scope, which focuses on identifying various forces as partial causes rather than calculating the consequences of many specific assumptions.

For this particular problem, the most detailed studies of the balance between selection and recombination have been made in classical Mendelian genetics. We may consider two interacting traits to be encoded by two genetic loci within a genome. Selection builds up a positive correlation between alleles that interact in a positively synergistic way. Recombination breaks apart those correlated allelic pairs.

Many studies have analyzed how selection and recombination interact to determine the allelic correlation between loci, typically measured as linkage disequilibrium \autocite{barton02the-effect,campos19evolution,hill68linkage,neher11genetic,}.

\subsection{Culture, transmissible environment}

I assumed two interacting species in a simple community. However, the role of correlations between traits is much broader. Any factor that affects the fitness of a focal individual and its transmission to the future may play a role.

For example, a culturally transmitted attribute can influence the fitness of various biological individuals and can often be partially cotransmitted with the trait values of some of the individuals in a community. Thus, selection and recombination between biological and cultural traits may happen in species that have cultural attributes \autocite{feldman96gene-culture,lehmann08the-coevolution}.

Similarly, an environmental attribute can influence fitness, with continuity of environmental states through time and space. If the environment is independent of the focal biological traits, then selection does not change the frequency of environments. However, it may be possible for selection to build up associations between trait values and environmental states \autocite{ravigne09live}.

Some studies focus on how organisms influence their environment, sometimes called niche construction \autocite{laland99evolutionary,odling-smee03niche}. Other studies focus on how the environment influences biological traits, phenotypic plasticity \autocite{west-eberhard03developmental,21phenotypic}. These cases differ from the simpler models above in which correlations arise primarily by a balance between selection and recombination. Ultimately, it would be useful to have a broad conceptual framework that included all of these examples as special cases of a general theory.

\section{Evolvability}

Evolvability describes the potential for a lineage to adapt to new challenges. Evolvability often depends on the common tradeoff between exploration and exploitation. Exploration enhances discovery of new solutions but reduces the efficiency of exploiting current solutions \autocite{payne19the-causes,wagner13robustness,earl04evolvability,kirschner98evolvability,pigliucci08is-evolvability,wagner96complex}.

In biology, a lineage might explore more widely by increasing the mechanisms that generate novel variation. Mutation, sexual reproduction, enhanced phenotypic flexibility, and learning create variability.

Extra variability and exploration often induce an immediate cost, degrading exploitation. Most mutations are deleterious. Sex requires various additional steps in reproduction and breaks up well-adapted genetic combinations \autocite{otto97deleterious,otto02resolving,otto09the-evolutionary}. Flexible phenotypes spend some time away from what has previously worked best.

With regard to temporal scale, exploitation is a continuous, immediate benefit. Exploration typically yields sporadic benefits at some future time. Spatially, the local benefits of exploitation are typically stronger than the distant benefits of exploration.

Evolvability usually depends on the interaction between two traits. One trait generates the variability of exploration. A variability-generating mechanism does not by itself solve an environmental challenge. Instead, it creates variability in a second trait that can adapt to novel challenge \autocite{karlin74towards,otto13evolution}.

For example, reduced DNA repair efficacy generates more mutants. But reduced repair does not directly do something that is useful, such as increase the uptake of food. Instead, it creates more variability in traits that influence how food is taken up. When the environment changes, a variant in food uptake may be just what is needed to cope with the new type of feeding challenge.

This distinction between generating variation and solving an environmental challenge means that we have a two-trait problem. As in the prior section, the correlation between the two traits plays a central role in evolutionary dynamics. Thus, for evolvability, we often have a problem that combines temporal scaling and the correlation between traits.

Here, I illustrate the key forces in the simplest way. Suppose fitness may be written as
\begin{equation*}
  w = (1-x)^n\, \tilde{x}^k,
\end{equation*}
in which $x$ is a trait that generates variation. An individual's fitness is reduced by $1-x$ in each time period because generating variation imposes a cost. Environmental challenges arise after $n$ time periods, with costs multiplying through the time periods.  

The term $\tilde{x}$ is the variability in a trait that can adapt to a novel environmental challenge, defined as
\begin{equation*}
  \tilde{x} = \Gr x + (1-\Gr)z^*,
\end{equation*}
in which $z^*$ is the population-wide value of $x$ at a candidate equilibrium, and $\Gr$ is the correlation within a lineage between the variance-generating trait, $x$, and the exploration-benefitting variability in that lineage, $\tilde{x}$. As in the prior section on multiple traits, the correlation between traits may be enhanced by selection and degraded by recombination or failure to cotransmit.

The calculation of $\Gr$ for the dynamics of a particular situation may be complex. Here, we invoke $\Gr$ as a force that influences evolvability in our context of explanation rather than calculation.

We can find a candidate optimum for $x$ by evaluating $\dd w/\dd x=0$ at $x=z^*$. That yields
\begin{equation*}
  z^*=\frac{k\Gr}{n+k\Gr}
\end{equation*}
This result is easier to parse as the ratio $z^*:1-z^*$, written as
\begin{equation*}
  k\Gr : n,
\end{equation*}
which is the ratio of the benefits of exploration by generating variation, $k\Gr$, relative to the benefits of exploitation by reducing the generation of variation, $n$.

We ignored reproductive value weightings in this example. If the population is growing, then natural selection favors earlier reproduction over later reproduction. That weighting of current time over future time favors greater exploitation and reduced evolvability. By contrast, a shrinking population favors later benefits over sooner ones, tending to enhance evolvability.

\section{Discussion}

Natural selection explains adaptation. For example, we say that mothers divide resources between sons and daughters to increase maternal success. Parasites compete within hosts to enhance their transmission. Transposons spread in genomes to raise their future representation in populations.

In each case, natural selection presumably tunes traits to improve reproductive success. But that perspective by itself also hides as much as it reveals.

A mother can outcompete a neighbor by making more sons and fewer daughters. Those sons gain a greater share of local mating, fathering more descendants. But with fewer daughters around, the additional sons compete for a smaller gain. The competitive mother gains against her neighbor but the group produces less grandprogeny. The local group loses against groups that make fewer sons and more daughters, competing less internally. Natural selection acts one way locally and another way globally.

Similarly, a parasite can gain an immediate benefit against its neighbor within a host by taking more resources. But too much exploitation can harm the host, slowly degrading the resource on which the parasite depends. Natural selection acts one way in the short term and another way in the long term.

A transposon's duplication to another genomic site enhances its short term success within its host. But too many transposons can harm that host's reproduction, which is also the long term reproduction of the transposons. Natural selection acts one way at the level of within host success and another way at the higher level of between host success.

Opposing forces arise in each case. At one scale, natural selection pushes in one direction. At another scale, natural selection pushes in another direction. Our initial idea of using natural selection to explain adaptation leads to a new question.

How do the opposing forces at different scales resolve? To understand that question, we run into two distinct approaches within the literature. The common theoretical approach is to make explicit assumptions about each aspect of a biological scenario and then work out how things change over time, the dynamics. That approach provides clarity about the relations between particular assumptions and outcomes.

Alternatively, one can leave out most of the details and analyze how a certain factor tends to push the system in a particular direction. For example, if we compare situations in which parasites in a host are more genetically similar versus less genetically similar, how does that difference alter the tendency to harm the host? This approach takes the all-else-equal analysis to its extreme, a kind of static analysis \autocite{frank22microbial}. How does a particular force act as a partial cause when considered in isolation?

These alternative approaches often seem as if they were at war with each other. The biology literature reflects the same tension described for economics by Samuelson \autocite{samuelson83foundations}
\begin{quote}
Often in the writings of economists the words “dynamic” and “static” are used as nothing more than synonyms for good and bad, realistic and unrealistic, simple and complex. We damn another man’s theory by terming it static, and advertise our own by calling it dynamic. Examples of this are too plentiful to require citation.
\end{quote}
The different perspectives are, of course, not actually in conflict. They simply provide different views of the same problem. I like to think of statics and dynamics as different tools, like a screwdriver and a hammer. It is good to bring both to a new job, you never know exactly what the challenge may be.

As Schumpeter pointed out in the quote given earlier, statics often leads the way. Because clear understanding of forces as partial causes provides the best explanations for realistic studies of complex natural phenomena.

For these reasons, I have focused in this article on providing a collection of primary opposing forces that act at different scales. Often in biology, interpreting and explaining traits depend on having this small catalog of forces, the entries to be applied as needed for each case.

I gave one extended example that compared statics and dynamics for the classic group selection problem. That example emphasized that one has to see problems from both perspectives to gain a complete view. And it also showed the benefit of simplicity, the gain achieved by statics when forming testable hypotheses. The static models typically suggest how altering a particular force changes a trait in a predictable way.

The remainder of the article filled in the catalog of models, each specifying simple comparative predictions. Overall, the focus on opposing forces at different temporal and spatial scales provided significant insight.

\section*{Acknowledgments}

\noindent Parts of this article were modified from earlier publications \autocite{frank98foundations,frank22microbial}. The Donald Bren Foundation and US National Science Foundation grant DEB-2325755 support my research.


\mybiblio	



\begin{thebibliography}{100}
\expandafter\ifx\csname url\endcsname\relax
  \def\url#1{\texttt{#1}}\fi
\expandafter\ifx\csname urlprefix\endcsname\relax\def\urlprefix{URL }\fi
\providecommand{\bibinfo}[2]{#2}
\providecommand{\eprint}[2][]{\url{#2}}

\bibitem{williams66adaptation}
\bibinfo{author}{Williams, G.~C.}
\newblock \emph{\bibinfo{title}{Adaptation and {N}atural {S}election}}
  (\bibinfo{publisher}{Princeton University Press},
  \bibinfo{address}{Princeton, NJ}, \bibinfo{year}{1966}).

\bibitem{hamilton75innate}
\bibinfo{author}{Hamilton, W.~D.}
\newblock \bibinfo{title}{Innate social aptitudes of man: an approach from
  evolutionary genetics}.
\newblock In \bibinfo{editor}{Fox, R.} (ed.)
  \emph{\bibinfo{booktitle}{{Biosocial Anthropology}}},
  \bibinfo{pages}{133--155} (\bibinfo{publisher}{Wiley}, \bibinfo{address}{New
  York}, \bibinfo{year}{1975}).

\bibitem{maynard-smith76group}
\bibinfo{author}{Maynard~Smith, J.}
\newblock \bibinfo{title}{Group selection}.
\newblock \emph{\bibinfo{journal}{Quarterly Review of Biology}}
  \textbf{\bibinfo{volume}{51}}, \bibinfo{pages}{277--283}
  (\bibinfo{year}{1976}).

\bibitem{wilson83the-group}
\bibinfo{author}{Wilson, D.~S.}
\newblock \bibinfo{title}{The group selection controversy: history and current
  status}.
\newblock \emph{\bibinfo{journal}{Annual Review of Ecology and Systematics}}
  \textbf{\bibinfo{volume}{14}}, \bibinfo{pages}{159--187}
  (\bibinfo{year}{1983}).

\bibitem{grafen84natural}
\bibinfo{author}{Grafen, A.}
\newblock \bibinfo{title}{Natural selection, kin selection and group
  selection}.
\newblock In \bibinfo{editor}{Krebs, J.~R.} \& \bibinfo{editor}{Davies, N.~B.}
  (eds.) \emph{\bibinfo{booktitle}{Behavioural Ecology}},
  \bibinfo{pages}{62--84} (\bibinfo{publisher}{Blackwell Scientific
  Publications}, \bibinfo{address}{Oxford}, \bibinfo{year}{1984}).

\bibitem{queller91group}
\bibinfo{author}{Queller, D.~C.}
\newblock \bibinfo{title}{Group selection and kin selection}.
\newblock \emph{\bibinfo{journal}{Trends in Ecology and Evolution}}
  \textbf{\bibinfo{volume}{6}}, \bibinfo{pages}{64} (\bibinfo{year}{1991}).

\bibitem{okasha06evolution}
\bibinfo{author}{Okasha, S.}
\newblock \emph{\bibinfo{title}{{Evolution and the Levels of Selection}}}
  (\bibinfo{publisher}{Oxford University Press}, \bibinfo{address}{New York},
  \bibinfo{year}{2006}).

\bibitem{west08social}
\bibinfo{author}{West, S.~A.}, \bibinfo{author}{Griffin, A.~S.} \&
  \bibinfo{author}{Gardner, A.}
\newblock \bibinfo{title}{Social semantics: how useful has group selection
  been?}
\newblock \emph{\bibinfo{journal}{Journal of Evolutionary Biology}}
  \textbf{\bibinfo{volume}{21}}, \bibinfo{pages}{374--385}
  (\bibinfo{year}{2008}).

\bibitem{gardner09capturing}
\bibinfo{author}{Gardner, A.} \& \bibinfo{author}{Grafen, A.}
\newblock \bibinfo{title}{Capturing the superorganism: a formal theory of group
  adaptation}.
\newblock \emph{\bibinfo{journal}{Journal of Evolutionary Biology}}
  \textbf{\bibinfo{volume}{22}}, \bibinfo{pages}{659--671}
  (\bibinfo{year}{2009}).

\bibitem{leigh10the-group}
\bibinfo{author}{Leigh, E.~G., Jr.}
\newblock \bibinfo{title}{The group selection controversy}.
\newblock \emph{\bibinfo{journal}{Journal of Evolutionary Biology}}
  \textbf{\bibinfo{volume}{23}}, \bibinfo{pages}{6--19} (\bibinfo{year}{2010}).

\bibitem{wilson75a-theory}
\bibinfo{author}{Wilson, D.~S.}
\newblock \bibinfo{title}{A theory of group selection}.
\newblock \emph{\bibinfo{journal}{Proceedings of the National Academy of
  Sciences USA}} \textbf{\bibinfo{volume}{72}}, \bibinfo{pages}{143--146}
  (\bibinfo{year}{1975}).

\bibitem{leigh77how-does}
\bibinfo{author}{Leigh, E.~G., Jr.}
\newblock \bibinfo{title}{How does selection reconcile individual advantage
  with the good of the group?}
\newblock \emph{\bibinfo{journal}{Proceedings of the National Academy of
  Sciences USA}} \textbf{\bibinfo{volume}{74}}, \bibinfo{pages}{4542--4546}
  (\bibinfo{year}{1977}).

\bibitem{frank98foundations}
\bibinfo{author}{Frank, S.~A.}
\newblock \emph{\bibinfo{title}{Foundations of {S}ocial {E}volution}}
  (\bibinfo{publisher}{Princeton University Press},
  \bibinfo{address}{Princeton, NJ}, \bibinfo{year}{1998}).

\bibitem{frank22microbial}
\bibinfo{author}{Frank, S.~A.}
\newblock \emph{\bibinfo{title}{{Microbial Life History: The Fundamental Forces
  of Biological Design}}} (\bibinfo{publisher}{Princeton University Press},
  \bibinfo{year}{2022}).

\bibitem{hardin68tragedy}
\bibinfo{author}{Hardin, G.}
\newblock \bibinfo{title}{{The tragedy of the commons}}.
\newblock \emph{\bibinfo{journal}{Science}} \textbf{\bibinfo{volume}{162}},
  \bibinfo{pages}{1243--1248} (\bibinfo{year}{1968}).

\bibitem{frank96models}
\bibinfo{author}{Frank, S.~A.}
\newblock \bibinfo{title}{Models of parasite virulence}.
\newblock \emph{\bibinfo{journal}{Quarterly Review of Biology}}
  \textbf{\bibinfo{volume}{71}}, \bibinfo{pages}{37--78}
  (\bibinfo{year}{1996}).

\bibitem{rankin07the-tragedy}
\bibinfo{author}{Rankin, D.~J.}, \bibinfo{author}{Bargum, K.} \&
  \bibinfo{author}{Kokko, H.}
\newblock \bibinfo{title}{{The tragedy of the commons in evolutionary
  biology}}.
\newblock \emph{\bibinfo{journal}{Trends Ecol. Evol.}}
  \textbf{\bibinfo{volume}{22}}, \bibinfo{pages}{643--651}
  (\bibinfo{year}{2007}).

\bibitem{alexander78group}
\bibinfo{author}{Alexander, R.~D.} \& \bibinfo{author}{Borgia, G.}
\newblock \bibinfo{title}{Group selection, altruism, and the levels of
  organization of life}.
\newblock \emph{\bibinfo{journal}{Annual Review of Ecology and Systematics}}
  \textbf{\bibinfo{volume}{9}}, \bibinfo{pages}{449--474}
  (\bibinfo{year}{1978}).

\bibitem{wade78a-critical}
\bibinfo{author}{Wade, M.~J.}
\newblock \bibinfo{title}{A critical review of the models of group selection}.
\newblock \emph{\bibinfo{journal}{Quarterly Review of Biology}}
  \textbf{\bibinfo{volume}{53}}, \bibinfo{pages}{101--114}
  (\bibinfo{year}{1978}).

\bibitem{wilson80the-natural}
\bibinfo{author}{Wilson, D.~S.}
\newblock \emph{\bibinfo{title}{The {N}atural {S}election of {P}opulations and
  {C}ommunities}} (\bibinfo{publisher}{Benjamin/Cummings},
  \bibinfo{address}{Menlo Park, CA}, \bibinfo{year}{1980}).

\bibitem{taylor96how-to-make}
\bibinfo{author}{Taylor, P.~D.} \& \bibinfo{author}{Frank, S.~A.}
\newblock \bibinfo{title}{How to make a kin selection model}.
\newblock \emph{\bibinfo{journal}{Journal of Theoretical Biology}}
  \textbf{\bibinfo{volume}{180}}, \bibinfo{pages}{27--37}
  (\bibinfo{year}{1996}).

\bibitem{frank94kin-selection}
\bibinfo{author}{Frank, S.~A.}
\newblock \bibinfo{title}{Kin selection and virulence in the evolution of
  protocells and parasites}.
\newblock \emph{\bibinfo{journal}{Proceedings of the Royal Society of London
  B}} \textbf{\bibinfo{volume}{258}}, \bibinfo{pages}{153--161}
  (\bibinfo{year}{1994}).

\bibitem{frank95mutual}
\bibinfo{author}{Frank, S.~A.}
\newblock \bibinfo{title}{Mutual policing and repression of competition in the
  evolution of cooperative groups}.
\newblock \emph{\bibinfo{journal}{Nature}} \textbf{\bibinfo{volume}{377}},
  \bibinfo{pages}{520--522} (\bibinfo{year}{1995}).

\bibitem{fisher58the-genetical}
\bibinfo{author}{Fisher, R.~A.}
\newblock \emph{\bibinfo{title}{The {G}enetical {T}heory of {N}atural
  {S}election}} (\bibinfo{publisher}{Dover Publications}, \bibinfo{address}{New
  York}, \bibinfo{year}{1958}), \bibinfo{edition}{2nd ed} edn.

\bibitem{price72fishers}
\bibinfo{author}{Price, G.~R.}
\newblock \bibinfo{title}{Fisher's `fundamental theorem' made clear}.
\newblock \emph{\bibinfo{journal}{Annals of Human Genetics}}
  \textbf{\bibinfo{volume}{36}}, \bibinfo{pages}{129--140}
  (\bibinfo{year}{1972}).

\bibitem{ewens89an-interpretation}
\bibinfo{author}{Ewens, W.~J.}
\newblock \bibinfo{title}{An interpretation and proof of the fundamental
  theorem of natural selection}.
\newblock \emph{\bibinfo{journal}{Theoretical Population Biology}}
  \textbf{\bibinfo{volume}{36}}, \bibinfo{pages}{167--180}
  (\bibinfo{year}{1989}).

\bibitem{frank92fishers}
\bibinfo{author}{Frank, S.~A.} \& \bibinfo{author}{Slatkin, M.}
\newblock \bibinfo{title}{Fisher's fundamental theorem of natural selection}.
\newblock \emph{\bibinfo{journal}{Trends in Ecology and Evolution}}
  \textbf{\bibinfo{volume}{7}}, \bibinfo{pages}{92--95} (\bibinfo{year}{1992}).

\bibitem{frank15dalemberts}
\bibinfo{author}{Frank, S.~A.}
\newblock \bibinfo{title}{D'{A}lembert's direct and inertial forces acting on
  populations: the {P}rice equation and the fundamental theorem of natural
  selection}.
\newblock \emph{\bibinfo{journal}{Entropy}} \textbf{\bibinfo{volume}{17}},
  \bibinfo{pages}{7087--7100} (\bibinfo{year}{2015}).

\bibitem{maynard-smith82evolution}
\bibinfo{author}{Maynard~Smith, J.}
\newblock \emph{\bibinfo{title}{Evolution and the {T}heory of {G}ames}}
  (\bibinfo{publisher}{Cambridge University Press},
  \bibinfo{address}{Cambridge}, \bibinfo{year}{1982}).

\bibitem{hamilton70selfish}
\bibinfo{author}{Hamilton, W.~D.}
\newblock \bibinfo{title}{Selfish and spiteful behaviour in an evolutionary
  model}.
\newblock \emph{\bibinfo{journal}{Nature}} \textbf{\bibinfo{volume}{228}},
  \bibinfo{pages}{1218--1220} (\bibinfo{year}{1970}).

\bibitem{queller92quantitative}
\bibinfo{author}{Queller, D.~C.}
\newblock \bibinfo{title}{Quantitative genetics, inclusive fitness, and group
  selection}.
\newblock \emph{\bibinfo{journal}{American Naturalist}}
  \textbf{\bibinfo{volume}{139}}, \bibinfo{pages}{540--558}
  (\bibinfo{year}{1992}).

\bibitem{frank86hierarchical}
\bibinfo{author}{Frank, S.~A.}
\newblock \bibinfo{title}{Hierarchical selection theory and sex ratios {I}.
  {G}eneral solutions for structured populations}.
\newblock \emph{\bibinfo{journal}{Theoretical Population Biology}}
  \textbf{\bibinfo{volume}{29}}, \bibinfo{pages}{312--342}
  (\bibinfo{year}{1986}).

\bibitem{eberhard80evolutionary}
\bibinfo{author}{Eberhard, W.~G.}
\newblock \bibinfo{title}{Evolutionary consequences of intracellular organelle
  competition}.
\newblock \emph{\bibinfo{journal}{Quarterly Review of Biology}}
  \textbf{\bibinfo{volume}{55}}, \bibinfo{pages}{231--249}
  (\bibinfo{year}{1980}).

\bibitem{cosmides81cytoplasmic}
\bibinfo{author}{Cosmides, L.~M.} \& \bibinfo{author}{Tooby, J.}
\newblock \bibinfo{title}{Cytoplasmic inheritance and intragenomic conflict}.
\newblock \emph{\bibinfo{journal}{Journal of Theoretical Biology}}
  \textbf{\bibinfo{volume}{89}}, \bibinfo{pages}{83--129}
  (\bibinfo{year}{1981}).

\bibitem{hoekstra87the-evolution}
\bibinfo{author}{Hoekstra, R.~F.}
\newblock \bibinfo{title}{The evolution of sexes}.
\newblock In \bibinfo{editor}{Stearns, S.~C.} (ed.)
  \emph{\bibinfo{booktitle}{The Evolution of Sex and its Consequences}},
  \bibinfo{pages}{59--91} (\bibinfo{publisher}{Birkhauser},
  \bibinfo{address}{Basel}, \bibinfo{year}{1987}).

\bibitem{hoekstra90evolution}
\bibinfo{author}{Hoekstra, R.~F.}
\newblock \bibinfo{title}{Evolution of uniparental inheritance of cyplasmic
  {D}{N}{A}}.
\newblock In \bibinfo{editor}{Maynard~Smith, J.} \& \bibinfo{editor}{Vida, G.}
  (eds.) \emph{\bibinfo{booktitle}{Organizational Constraints on the Dynamics
  of Evolution}}, \bibinfo{pages}{269--280} (\bibinfo{publisher}{Manchester
  University Press}, \bibinfo{address}{Manchester}, \bibinfo{year}{1990}).

\bibitem{frank96host-symbiont}
\bibinfo{author}{Frank, S.~A.}
\newblock \bibinfo{title}{Host-symbiont conflict over the mixing of symbiotic
  lineages}.
\newblock \emph{\bibinfo{journal}{Proceedings of the Royal Society of London
  B}} \textbf{\bibinfo{volume}{263}}, \bibinfo{pages}{339--344}
  (\bibinfo{year}{1996}).

\bibitem{ewald94the-evolution}
\bibinfo{author}{Ewald, P.~W.}
\newblock \emph{\bibinfo{title}{The {E}volution of {I}nfectious {D}isease}}
  (\bibinfo{publisher}{Oxford University Press}, \bibinfo{address}{Oxford},
  \bibinfo{year}{1994}).

\bibitem{szathmary87group}
\bibinfo{author}{Szathmáry, E.} \& \bibinfo{author}{Demeter, L.}
\newblock \bibinfo{title}{Group selection of early replicators and the origin
  of life}.
\newblock \emph{\bibinfo{journal}{J. Theor. Biol.}}
  \textbf{\bibinfo{volume}{128}}, \bibinfo{pages}{463--86}
  (\bibinfo{year}{1987}).

\bibitem{levin81selection}
\bibinfo{author}{Levin, S.~A.} \& \bibinfo{author}{Pimental, D.}
\newblock \bibinfo{title}{Selection of intermediate rates of increase in
  parasite-host systems}.
\newblock \emph{\bibinfo{journal}{American Naturalist}}
  \textbf{\bibinfo{volume}{117}}, \bibinfo{pages}{308--315}
  (\bibinfo{year}{1981}).

\bibitem{anderson82coevolution}
\bibinfo{author}{Anderson, R.~M.} \& \bibinfo{author}{May, R.~M.}
\newblock \bibinfo{title}{Coevolution of hosts and parasites}.
\newblock \emph{\bibinfo{journal}{Parasitology}} \textbf{\bibinfo{volume}{85}},
  \bibinfo{pages}{411--426} (\bibinfo{year}{1982}).

\bibitem{ewald83host-parasite}
\bibinfo{author}{Ewald, P.~W.}
\newblock \bibinfo{title}{Host-parasite relations, vectors, and the evolution
  of disease severity}.
\newblock \emph{\bibinfo{journal}{Annual Review of Ecology and Systematics}}
  \textbf{\bibinfo{volume}{14}}, \bibinfo{pages}{465--485}
  (\bibinfo{year}{1983}).

\bibitem{bremermann83a-game-theoretical}
\bibinfo{author}{Bremermann, H.~J.} \& \bibinfo{author}{Pickering, J.}
\newblock \bibinfo{title}{A game-theoretical model of parasite virulence}.
\newblock \emph{\bibinfo{journal}{Journal of Theoretical Biology}}
  \textbf{\bibinfo{volume}{100}}, \bibinfo{pages}{411--426}
  (\bibinfo{year}{1983}).

\bibitem{hamilton67extraordinary}
\bibinfo{author}{Hamilton, W.~D.}
\newblock \bibinfo{title}{Extraordinary sex ratios}.
\newblock \emph{\bibinfo{journal}{Science}} \textbf{\bibinfo{volume}{156}},
  \bibinfo{pages}{477--488} (\bibinfo{year}{1967}).

\bibitem{colwell81group}
\bibinfo{author}{Colwell, R.~K.}
\newblock \bibinfo{title}{Group selection is implicated in the evolution of
  female-biased sex ratios}.
\newblock \emph{\bibinfo{journal}{Nature}} \textbf{\bibinfo{volume}{290}},
  \bibinfo{pages}{401--404} (\bibinfo{year}{1981}).

\bibitem{alexander74the-evolution}
\bibinfo{author}{Alexander, R.~D.}
\newblock \bibinfo{title}{The evolution of social behavior}.
\newblock \emph{\bibinfo{journal}{Annual Review of Ecology and Systematics}}
  \textbf{\bibinfo{volume}{5}}, \bibinfo{pages}{325--383}
  (\bibinfo{year}{1974}).

\bibitem{clark78sex-ratio}
\bibinfo{author}{Clark, A.~B.}
\newblock \bibinfo{title}{Sex ratio and local resource competition in a
  prosiminian primate}.
\newblock \emph{\bibinfo{journal}{Science}} \textbf{\bibinfo{volume}{201}},
  \bibinfo{pages}{163--165} (\bibinfo{year}{1978}).

\bibitem{frank86the-genetic}
\bibinfo{author}{Frank, S.~A.}
\newblock \bibinfo{title}{The genetic value of sons and daughters}.
\newblock \emph{\bibinfo{journal}{Heredity}} \textbf{\bibinfo{volume}{56}},
  \bibinfo{pages}{351--354} (\bibinfo{year}{1986}).

\bibitem{taylor92altruism}
\bibinfo{author}{Taylor, P.~D.}
\newblock \bibinfo{title}{Altruism in viscous populations---an inclusive
  fitness approach}.
\newblock \emph{\bibinfo{journal}{Evolutionary Ecology}}
  \textbf{\bibinfo{volume}{6}}, \bibinfo{pages}{352--356}
  (\bibinfo{year}{1992}).

\bibitem{wilson92can-altruism}
\bibinfo{author}{Wilson, D.~S.}, \bibinfo{author}{Pollock, G.~B.} \&
  \bibinfo{author}{Dugatkin, L.~A.}
\newblock \bibinfo{title}{Can altruism evolve in purely viscous populations?}
\newblock \emph{\bibinfo{journal}{Evolutionary Ecology}}
  \textbf{\bibinfo{volume}{6}}, \bibinfo{pages}{331--341}
  (\bibinfo{year}{1992}).

\bibitem{hamilton77dispersal}
\bibinfo{author}{Hamilton, W.~D.} \& \bibinfo{author}{May, R.~M.}
\newblock \bibinfo{title}{Dispersal in stable habitats}.
\newblock \emph{\bibinfo{journal}{Nature}} \textbf{\bibinfo{volume}{269}},
  \bibinfo{pages}{578--581} (\bibinfo{year}{1977}).

\bibitem{ronce07how-does}
\bibinfo{author}{Ronce, O.}
\newblock \bibinfo{title}{How does it feel to be like a rolling stone? ten
  questions about dispersal evolution}.
\newblock \emph{\bibinfo{journal}{Annual Review of Ecology and Systematics}}
  \textbf{\bibinfo{volume}{38}}, \bibinfo{pages}{231--253}
  (\bibinfo{year}{2007}).

\bibitem{frank86dispersal}
\bibinfo{author}{Frank, S.~A.}
\newblock \bibinfo{title}{Dispersal polymorphisms in subdivided populations}.
\newblock \emph{\bibinfo{journal}{Journal of Theoretical Biology}}
  \textbf{\bibinfo{volume}{122}}, \bibinfo{pages}{303--309}
  (\bibinfo{year}{1986}).

\bibitem{karlin74towards}
\bibinfo{author}{Karlin, S.} \& \bibinfo{author}{McGregor, J.}
\newblock \bibinfo{title}{Towards a theory of the evolution of modifier genes}.
\newblock \emph{\bibinfo{journal}{Theoretical Population Biology}}
  \textbf{\bibinfo{volume}{5}}, \bibinfo{pages}{59--103}
  (\bibinfo{year}{1974}).

\bibitem{otto13evolution}
\bibinfo{author}{Otto, S.~P.}, \bibinfo{author}{Baum, D.} \&
  \bibinfo{author}{Futuyma, D.}
\newblock \bibinfo{title}{Evolution of modifier genes and biological systems}.
\newblock In \bibinfo{editor}{Losos, J.~B.} (ed.) \emph{\bibinfo{booktitle}{The
  Princeton Guide to Evolution}}, \bibinfo{pages}{255--262}
  (\bibinfo{publisher}{Princeton University Press},
  \bibinfo{address}{Princeton, NJ}, \bibinfo{year}{2013}).

\bibitem{eshel96on-the-changing}
\bibinfo{author}{Eshel, I.}
\newblock \bibinfo{title}{On the changing concept of evolutionary population
  stability as a reflection of changing point of view in the quantitative
  theory of evolution}.
\newblock \emph{\bibinfo{journal}{Journal of Mathematical Biology}}
  \textbf{\bibinfo{volume}{34}}, \bibinfo{pages}{485--510}
  (\bibinfo{year}{1996}).

\bibitem{diekmann04a-beginners}
\bibinfo{author}{Diekmann, O.} \emph{et~al.}
\newblock \bibinfo{title}{A beginner's guide to adaptive dynamics}.
\newblock \emph{\bibinfo{journal}{Banach Center Publications}}
  \textbf{\bibinfo{volume}{63}}, \bibinfo{pages}{47--86}
  (\bibinfo{year}{2004}).

\bibitem{mcgill07evolutionary}
\bibinfo{author}{McGill, B.~J.} \& \bibinfo{author}{Brown, J.~S.}
\newblock \bibinfo{title}{Evolutionary game theory and adaptive dynamics of
  continuous traits}.
\newblock \emph{\bibinfo{journal}{Annu. Rev. Ecol. Evol. Syst.}}
  \textbf{\bibinfo{volume}{38}}, \bibinfo{pages}{403--435}
  (\bibinfo{year}{2007}).

\bibitem{hibbeler16engineering}
\bibinfo{author}{Hibbeler, R.~C.}
\newblock \emph{\bibinfo{title}{Engineering Mechanics: Statics}}
  (\bibinfo{publisher}{Pearson}, \bibinfo{address}{Hoboken, NJ},
  \bibinfo{year}{2016}), \bibinfo{edition}{14} edn.

\bibitem{mankiw24principles}
\bibinfo{author}{Mankiw, N.~G.}
\newblock \emph{\bibinfo{title}{Principles of Economics}}
  (\bibinfo{publisher}{Cengage Learning}, \bibinfo{address}{Boston, MA},
  \bibinfo{year}{2024}), \bibinfo{edition}{10} edn.

\bibitem{samuelson83foundations}
\bibinfo{author}{Samuelson, P.~A.}
\newblock \emph{\bibinfo{title}{Foundations of {E}conomic {A}nalysis, Enlarged
  ed}} (\bibinfo{publisher}{Harvard University Press},
  \bibinfo{address}{Cambridge, MA}, \bibinfo{year}{1983}).

\bibitem{spencer05adaptive}
\bibinfo{author}{Spencer, H.} \& \bibinfo{author}{Feldman, M.}
\newblock \bibinfo{title}{Adaptive dynamics, game theory and evolutionary
  population genetics}.
\newblock \emph{\bibinfo{journal}{Journal of Evolutionary Biology}}
  \textbf{\bibinfo{volume}{18}}, \bibinfo{pages}{1191--1193}
  (\bibinfo{year}{2005}).

\bibitem{lanczos86the-variational}
\bibinfo{author}{Lanczos, C.}
\newblock \emph{\bibinfo{title}{The Variational Principles of Mechanics}}
  (\bibinfo{publisher}{Dover Publications}, \bibinfo{address}{New York},
  \bibinfo{year}{1986}), \bibinfo{edition}{4th ed} edn.

\bibitem{silberberg74a-revision}
\bibinfo{author}{Silberberg, E.}
\newblock \bibinfo{title}{A revision of comparative statics methodology in
  economics, or, how to do comparative statics on the back of an envelope}.
\newblock \emph{\bibinfo{journal}{Journal of Economic Theory}}
  \textbf{\bibinfo{volume}{7}}, \bibinfo{pages}{159--172}
  (\bibinfo{year}{1974}).

\bibitem{milgrom94monotone}
\bibinfo{author}{Milgrom, P.} \& \bibinfo{author}{Shannon, C.}
\newblock \bibinfo{title}{Monotone comparative statics}.
\newblock \emph{\bibinfo{journal}{Econometrica}} \textbf{\bibinfo{volume}{62}},
  \bibinfo{pages}{157--180} (\bibinfo{year}{1994}).

\bibitem{varian92microeconomic}
\bibinfo{author}{Varian, H.~R.}
\newblock \emph{\bibinfo{title}{Microeconomic Analysis}}
  (\bibinfo{publisher}{W. W. Norton \& Company}, \bibinfo{address}{New York},
  \bibinfo{year}{1992}), \bibinfo{edition}{3rd} edn.

\bibitem{judd82an-alternative}
\bibinfo{author}{Judd, K.~L.}
\newblock \bibinfo{title}{An alternative to steady-state comparisons in perfect
  foresight models}.
\newblock \emph{\bibinfo{journal}{Economics Letters}}
  \textbf{\bibinfo{volume}{10}}, \bibinfo{pages}{55--59}
  (\bibinfo{year}{1982}).

\bibitem{chiang92elements}
\bibinfo{author}{Chiang, A.~C.}
\newblock \emph{\bibinfo{title}{Elements of Dynamic Optimization}}
  (\bibinfo{publisher}{McGraw-Hill}, \bibinfo{address}{New York},
  \bibinfo{year}{1992}).

\bibitem{kamien91dynamic}
\bibinfo{author}{Kamien, M.~I.} \& \bibinfo{author}{Schwartz, N.~L.}
\newblock \emph{\bibinfo{title}{Dynamic Optimization: The Calculus of
  Variations and Optimal Control in Economics and Management}}
  (\bibinfo{publisher}{Elsevier}, \bibinfo{address}{Amsterdam},
  \bibinfo{year}{1991}), \bibinfo{edition}{3rd} edn.

\bibitem{frank24natural}
\bibinfo{author}{Frank, S.~A.}
\newblock \bibinfo{title}{Natural selection at multiple scales: {Julia}
  software code} (\bibinfo{year}{2024}).
\newblock
  \urlprefix\url{https://github.com/evolbio/NatSel_MultiScale/releases/tag/Version_1.0.1}.

\bibitem{levin94short-sighted}
\bibinfo{author}{Levin, B.~R.} \& \bibinfo{author}{Bull, J.~J.}
\newblock \bibinfo{title}{Short-sighted evolution and the virulence of
  pathogenic microorganisms}.
\newblock \emph{\bibinfo{journal}{Trends in Microbiology}}
  \textbf{\bibinfo{volume}{2}}, \bibinfo{pages}{76--81} (\bibinfo{year}{1994}).

\bibitem{frank10the-trade-off}
\bibinfo{author}{Frank, S.~A.}
\newblock \bibinfo{title}{The trade-off between rate and yield in the design of
  microbial metabolism}.
\newblock \emph{\bibinfo{journal}{Journal of Evolutionary Biology}}
  \textbf{\bibinfo{volume}{23}}, \bibinfo{pages}{609--613}
  (\bibinfo{year}{2010}).

\bibitem{frank10microbial}
\bibinfo{author}{Frank, S.~A.}
\newblock \bibinfo{title}{Microbial secretor--cheater dynamics}.
\newblock \emph{\bibinfo{journal}{Philosophical Transactions of the Royal
  Society B}} \textbf{\bibinfo{volume}{365}}, \bibinfo{pages}{2515--2522}
  (\bibinfo{year}{2010}).

\bibitem{frank13microbial}
\bibinfo{author}{Frank, S.~A.}
\newblock \bibinfo{title}{Microbial evolution: regulatory design prevents
  cancer-like overgrowths}.
\newblock \emph{\bibinfo{journal}{Current Biology}}
  \textbf{\bibinfo{volume}{23}}, \bibinfo{pages}{R343--R346}
  (\bibinfo{year}{2013}).

\bibitem{schumpeter54history}
\bibinfo{author}{Schumpeter, J.~A.}
\newblock \emph{\bibinfo{title}{History of {E}conomic {A}nalysis}}
  (\bibinfo{publisher}{Oxford University Press}, \bibinfo{address}{New York},
  \bibinfo{year}{1954}).

\bibitem{dirac58the-principles}
\bibinfo{author}{Dirac, P. A.~M.}
\newblock \emph{\bibinfo{title}{The Principles of Quantum Mechanics}}
  (\bibinfo{publisher}{Oxford University Press}, \bibinfo{address}{Oxford},
  \bibinfo{year}{1958}), \bibinfo{edition}{4th} edn.

\bibitem{feynman67the-character}
\bibinfo{author}{Feynman, R.~P.}
\newblock \emph{\bibinfo{title}{The Character of Physical Law}}
  (\bibinfo{publisher}{MIT Press}, \bibinfo{address}{Cambridge, MA},
  \bibinfo{year}{1967}).

\bibitem{fuchs00quantum}
\bibinfo{author}{Fuchs, C.~A.} \& \bibinfo{author}{Peres, A.}
\newblock \bibinfo{title}{Quantum theory needs no ``interpretation''}.
\newblock \emph{\bibinfo{journal}{Physics Today}}
  \textbf{\bibinfo{volume}{53}}, \bibinfo{pages}{70--71}
  (\bibinfo{year}{2000}).

\bibitem{mermin89whats}
\bibinfo{author}{Mermin, N.~D.}
\newblock \bibinfo{title}{What's wrong with this pillow?}
\newblock \emph{\bibinfo{journal}{Physics Today}}
  \textbf{\bibinfo{volume}{42}}, \bibinfo{pages}{9--11} (\bibinfo{year}{1989}).

\bibitem{bell87speakable}
\bibinfo{author}{Bell, J.~S.}
\newblock \emph{\bibinfo{title}{Speakable and Unspeakable in Quantum
  Mechanics}} (\bibinfo{publisher}{Cambridge University Press},
  \bibinfo{address}{Cambridge}, \bibinfo{year}{1987}).

\bibitem{bohr49discussion}
\bibinfo{author}{Bohr, N.}
\newblock \bibinfo{title}{Discussion with {E}instein on epistemological
  problems in atomic physics}.
\newblock In \bibinfo{editor}{Schilpp, P.~A.} (ed.)
  \emph{\bibinfo{booktitle}{Albert Einstein: Philosopher-Scientist}},
  \bibinfo{pages}{201--241} (\bibinfo{publisher}{Open Court Publishing},
  \bibinfo{address}{La Salle, IL}, \bibinfo{year}{1949}).

\bibitem{einstein35can-quantum-mechanical}
\bibinfo{author}{Einstein, A.}, \bibinfo{author}{Podolsky, B.} \&
  \bibinfo{author}{Rosen, N.}
\newblock \bibinfo{title}{Can quantum-mechanical description of physical
  reality be considered complete?}
\newblock \emph{\bibinfo{journal}{Physical Review}}
  \textbf{\bibinfo{volume}{47}}, \bibinfo{pages}{777--780}
  (\bibinfo{year}{1935}).

\bibitem{heisenberg58physics}
\bibinfo{author}{Heisenberg, W.}
\newblock \emph{\bibinfo{title}{Physics and Philosophy: The Revolution in
  Modern Science}} (\bibinfo{publisher}{Harper \& Row}, \bibinfo{address}{New
  York}, \bibinfo{year}{1958}).

\bibitem{bohm80wholeness}
\bibinfo{author}{Bohm, D.}
\newblock \emph{\bibinfo{title}{Wholeness and the Implicate Order}}
  (\bibinfo{publisher}{Routledge}, \bibinfo{address}{London},
  \bibinfo{year}{1980}).

\bibitem{bricmont16making}
\bibinfo{author}{Bricmont, J.}
\newblock \emph{\bibinfo{title}{Making Sense of Quantum Mechanics}}
  (\bibinfo{publisher}{Springer}, \bibinfo{address}{Cham},
  \bibinfo{year}{2016}).

\bibitem{callebaut12scientific}
\bibinfo{author}{Callebaut, W.}
\newblock \bibinfo{title}{Scientific perspectivism: A philosopher of science's
  response to the challenge of big data biology}.
\newblock \emph{\bibinfo{journal}{Studies in History and Philosophy of Science
  Part C: Studies in History and Philosophy of Biological and Biomedical
  Sciences}} \textbf{\bibinfo{volume}{43}}, \bibinfo{pages}{69--80}
  (\bibinfo{year}{2012}).

\bibitem{mayr82the-growth}
\bibinfo{author}{Mayr, E.}
\newblock \emph{\bibinfo{title}{The Growth of Biological Thought: Diversity,
  Evolution, and Inheritance}} (\bibinfo{publisher}{Harvard University Press},
  \bibinfo{address}{Cambridge, MA}, \bibinfo{year}{1982}).

\bibitem{mccloskey98the-rhetoric}
\bibinfo{author}{McCloskey, D.~N.}
\newblock \emph{\bibinfo{title}{The Rhetoric of Economics}}
  (\bibinfo{publisher}{University of Wisconsin Press},
  \bibinfo{address}{Madison}, \bibinfo{year}{1998}), \bibinfo{edition}{2} edn.

\bibitem{rodrik15economics}
\bibinfo{author}{Rodrik, D.}
\newblock \emph{\bibinfo{title}{Economics Rules: The Rights and Wrongs of the
  Dismal Science}} (\bibinfo{publisher}{W. W. Norton \& Company},
  \bibinfo{address}{New York}, \bibinfo{year}{2015}).

\bibitem{jevons71the-theory}
\bibinfo{author}{Jevons, W.~S.}
\newblock \emph{\bibinfo{title}{The Theory of Political Economy}}
  (\bibinfo{publisher}{Macmillan and Co.}, \bibinfo{address}{London},
  \bibinfo{year}{1871}).

\bibitem{marshall90principles}
\bibinfo{author}{Marshall, A.}
\newblock \emph{\bibinfo{title}{Principles of Economics}}
  (\bibinfo{publisher}{Macmillan}, \bibinfo{address}{London},
  \bibinfo{year}{1890}).
\newblock \bibinfo{note}{First Edition}.

\bibitem{mas-colell95microeconomic}
\bibinfo{author}{Mas-Colell, A.}, \bibinfo{author}{Whinston, M.~D.} \&
  \bibinfo{author}{Green, J.~R.}
\newblock \emph{\bibinfo{title}{Microeconomic Theory}}
  (\bibinfo{publisher}{Oxford University Press}, \bibinfo{address}{New York},
  \bibinfo{year}{1995}).

\bibitem{blaug97economic}
\bibinfo{author}{Blaug, M.}
\newblock \emph{\bibinfo{title}{Economic Theory in Retrospect}}
  (\bibinfo{publisher}{Cambridge University Press},
  \bibinfo{address}{Cambridge}, \bibinfo{year}{1997}), \bibinfo{edition}{5}
  edn.

\bibitem{robertson66a-mathematical}
\bibinfo{author}{Robertson, A.}
\newblock \bibinfo{title}{A mathematical model of the culling process in dairy
  cattle}.
\newblock \emph{\bibinfo{journal}{Animal Production}}
  \textbf{\bibinfo{volume}{8}}, \bibinfo{pages}{95--108}
  (\bibinfo{year}{1966}).

\bibitem{price70selection}
\bibinfo{author}{Price, G.~R.}
\newblock \bibinfo{title}{Selection and covariance}.
\newblock \emph{\bibinfo{journal}{Nature}} \textbf{\bibinfo{volume}{227}},
  \bibinfo{pages}{520--521} (\bibinfo{year}{1970}).

\bibitem{price72extension}
\bibinfo{author}{Price, G.~R.}
\newblock \bibinfo{title}{Extension of covariance selection mathematics}.
\newblock \emph{\bibinfo{journal}{Annals of Human Genetics}}
  \textbf{\bibinfo{volume}{35}}, \bibinfo{pages}{485--490}
  (\bibinfo{year}{1972}).

\bibitem{frank97the-price}
\bibinfo{author}{Frank, S.~A.}
\newblock \bibinfo{title}{The {{P}}rice equation, {{F}}isher's fundamental
  theorem, kin selection, and causal analysis}.
\newblock \emph{\bibinfo{journal}{Evolution}} \textbf{\bibinfo{volume}{51}},
  \bibinfo{pages}{1712--1729} (\bibinfo{year}{1997}).

\bibitem{wolf99interacting}
\bibinfo{author}{Wolf, J.~B.}, \bibinfo{author}{Brodie~III, E.~D.} \&
  \bibinfo{author}{Moore, A.~J.}
\newblock \bibinfo{title}{Interacting phenotypes and the evolutionary process.
  {II}. {S}election resulting from social interactions}.
\newblock \emph{\bibinfo{journal}{American Naturalist}}
  \textbf{\bibinfo{volume}{153}}, \bibinfo{pages}{254--266}
  (\bibinfo{year}{1999}).

\bibitem{frank12naturalb}
\bibinfo{author}{Frank, S.~A.}
\newblock \bibinfo{title}{Natural selection. {IV}. {T}he {P}rice equation}.
\newblock \emph{\bibinfo{journal}{Journal of Evolutionary Biology}}
  \textbf{\bibinfo{volume}{25}}, \bibinfo{pages}{1002--1019}
  (\bibinfo{year}{2012}).

\bibitem{frank97multivariate}
\bibinfo{author}{Frank, S.~A.}
\newblock \bibinfo{title}{Multivariate analysis of correlated selection and kin
  selection, with an {E}{S}{S} maximization method}.
\newblock \emph{\bibinfo{journal}{Journal of Theoretical Biology}}
  \textbf{\bibinfo{volume}{189}}, \bibinfo{pages}{307--316}
  (\bibinfo{year}{1997}).

\bibitem{frank94genetics}
\bibinfo{author}{Frank, S.~A.}
\newblock \bibinfo{title}{Genetics of mutualism: the evolution of altruism
  between species}.
\newblock \emph{\bibinfo{journal}{Journal of Theoretical Biology}}
  \textbf{\bibinfo{volume}{170}}, \bibinfo{pages}{393--400}
  (\bibinfo{year}{1994}).

\bibitem{aumann74subjectivity}
\bibinfo{author}{Aumann, R.~J.}
\newblock \bibinfo{title}{Subjectivity and correlation in randomized
  strategies}.
\newblock \emph{\bibinfo{journal}{Journal of Mathematical Economics}}
  \textbf{\bibinfo{volume}{1}}, \bibinfo{pages}{67--96} (\bibinfo{year}{1974}).

\bibitem{skyrms96evolution}
\bibinfo{author}{Skyrms, B.}
\newblock \emph{\bibinfo{title}{Evolution of the {S}ocial {C}ontract}}
  (\bibinfo{publisher}{Cambridge University Press},
  \bibinfo{address}{Cambridge}, \bibinfo{year}{1996}).

\bibitem{moore97interacting}
\bibinfo{author}{Moore, A.~J.}, \bibinfo{author}{Brodie, E.~D.} \&
  \bibinfo{author}{Wolf, J.~B.}
\newblock \bibinfo{title}{Interacting phenotypes and the evolutionary process:
  {I}. {D}irect and indirect effects of social interactions}.
\newblock \emph{\bibinfo{journal}{Evolution}} \textbf{\bibinfo{volume}{51}},
  \bibinfo{pages}{1352--1362} (\bibinfo{year}{1997}).

\bibitem{taylor07direct}
\bibinfo{author}{Taylor, P.~D.}, \bibinfo{author}{Wild, G.} \&
  \bibinfo{author}{Gardner, A.}
\newblock \bibinfo{title}{Direct fitness or inclusive fitness: how shall we
  model kin selection?}
\newblock \emph{\bibinfo{journal}{Journal of Evolutionary Biology}}
  \textbf{\bibinfo{volume}{20}}, \bibinfo{pages}{301--309}
  (\bibinfo{year}{2007}).

\bibitem{gardner11the-genetical}
\bibinfo{author}{Gardner, A.}, \bibinfo{author}{West, S.~A.} \&
  \bibinfo{author}{Wild, G.}
\newblock \bibinfo{title}{The genetical theory of kin selection}.
\newblock \emph{\bibinfo{journal}{Journal of Evolutionary Biology}}
  \textbf{\bibinfo{volume}{24}}, \bibinfo{pages}{1020--1043}
  (\bibinfo{year}{2011}).

\bibitem{mcglothlin10interacting}
\bibinfo{author}{McGlothlin, J.~W.}, \bibinfo{author}{Moore, A.~J.},
  \bibinfo{author}{Wolf, J.~B.} \& \bibinfo{author}{Brodie~III, E.~D.}
\newblock \bibinfo{title}{Interacting phenotypes and the evolutionary process.
  {III}. {S}ocial evolution}.
\newblock \emph{\bibinfo{journal}{Evolution}} \textbf{\bibinfo{volume}{64}},
  \bibinfo{pages}{2558--2574} (\bibinfo{year}{2010}).

\bibitem{queller11expanded}
\bibinfo{author}{Queller, D.~C.}
\newblock \bibinfo{title}{Expanded social fitness and hamilton's rule for kin,
  kith, and kind}.
\newblock \emph{\bibinfo{journal}{Proceedings of the National Academy of
  Sciences}} \textbf{\bibinfo{volume}{108}}, \bibinfo{pages}{10792--10799}
  (\bibinfo{year}{2011}).

\bibitem{taylor90allele-frequency}
\bibinfo{author}{Taylor, P.~D.}
\newblock \bibinfo{title}{Allele-frequency change in a class-structured
  population}.
\newblock \emph{\bibinfo{journal}{American Naturalist}}
  \textbf{\bibinfo{volume}{135}}, \bibinfo{pages}{95--106}
  (\bibinfo{year}{1990}).

\bibitem{charlesworth94evolution}
\bibinfo{author}{Charlesworth, B.}
\newblock \emph{\bibinfo{title}{Evolution in {A}ge--{S}tructured
  {P}opulations}} (\bibinfo{publisher}{Cambridge University Press},
  \bibinfo{address}{Cambridge}, \bibinfo{year}{1994}), \bibinfo{edition}{2nd
  ed} edn.

\bibitem{frank10demography}
\bibinfo{author}{Frank, S.~A.}
\newblock \bibinfo{title}{Demography and the tragedy of the commons}.
\newblock \emph{\bibinfo{journal}{Journal of Evolutionary Biology}}
  \textbf{\bibinfo{volume}{23}}, \bibinfo{pages}{32--39}
  (\bibinfo{year}{2010}).

\bibitem{foster06a-general}
\bibinfo{author}{Foster, K.~R.} \& \bibinfo{author}{Wenseleers, T.}
\newblock \bibinfo{title}{A general model for the evolution of mutualisms}.
\newblock \emph{\bibinfo{journal}{Journal of Evolutionary Biology}}
  \textbf{\bibinfo{volume}{19}}, \bibinfo{pages}{1283--1293}
  (\bibinfo{year}{2006}).

\bibitem{leigh-jr10the-evolution}
\bibinfo{author}{Leigh~Jr, E.~G.}
\newblock \bibinfo{title}{{The evolution of mutualism}}.
\newblock \emph{\bibinfo{journal}{Journal of Evolutionary Biology}}
  \textbf{\bibinfo{volume}{23}}, \bibinfo{pages}{2507--2528}
  (\bibinfo{year}{2010}).
\newblock \urlprefix\url{https://doi.org/10.1111/j.1420-9101.2010.02114.x}.

\bibitem{sachs04the-evolution}
\bibinfo{author}{Sachs, J.~L.}, \bibinfo{author}{Mueller, U.~G.},
  \bibinfo{author}{Wilcox, T.~P.} \& \bibinfo{author}{Bull, J.~J.}
\newblock \bibinfo{title}{The evolution of cooperation}.
\newblock \emph{\bibinfo{journal}{Quarterly Review of Biology}}
  \textbf{\bibinfo{volume}{79}}, \bibinfo{pages}{135--160}
  (\bibinfo{year}{2004}).

\bibitem{frank95the-origin}
\bibinfo{author}{Frank, S.~A.}
\newblock \bibinfo{title}{The origin of synergistic symbiosis}.
\newblock \emph{\bibinfo{journal}{Journal of Theoretical Biology}}
  \textbf{\bibinfo{volume}{176}}, \bibinfo{pages}{403--410}
  (\bibinfo{year}{1995}).

\bibitem{frank97models}
\bibinfo{author}{Frank, S.~A.}
\newblock \bibinfo{title}{Models of symbiosis}.
\newblock \emph{\bibinfo{journal}{American Naturalist}}
  \textbf{\bibinfo{volume}{150}}, \bibinfo{pages}{S80--S99}
  (\bibinfo{year}{1997}).

\bibitem{barton02the-effect}
\bibinfo{author}{Barton, N.~H.} \& \bibinfo{author}{Otto, S.~P.}
\newblock \bibinfo{title}{The effect of linkage on limits to artificial
  selection}.
\newblock \emph{\bibinfo{journal}{Genetics}} \textbf{\bibinfo{volume}{161}},
  \bibinfo{pages}{1605--1619} (\bibinfo{year}{2002}).

\bibitem{campos19evolution}
\bibinfo{author}{Campos, J.~L.} \& \bibinfo{author}{Charlesworth, B.}
\newblock \bibinfo{title}{Evolution of the linkage architecture of complex
  traits}.
\newblock \emph{\bibinfo{journal}{Nature Reviews Genetics}}
  \textbf{\bibinfo{volume}{20}}, \bibinfo{pages}{483--493}
  (\bibinfo{year}{2019}).

\bibitem{hill68linkage}
\bibinfo{author}{Hill, W.~G.} \& \bibinfo{author}{Robertson, A.}
\newblock \bibinfo{title}{Linkage disequilibrium in finite populations}.
\newblock \emph{\bibinfo{journal}{Theoretical and Applied Genetics}}
  \textbf{\bibinfo{volume}{38}}, \bibinfo{pages}{226--231}
  (\bibinfo{year}{1968}).

\bibitem{neher11genetic}
\bibinfo{author}{Neher, R.~A.}
\newblock \bibinfo{title}{Genetic draft, selective interference, and population
  genetics of rapid adaptation}.
\newblock \emph{\bibinfo{journal}{Annual Review of Ecology, Evolution, and
  Systematics}} \textbf{\bibinfo{volume}{42}}, \bibinfo{pages}{287--311}
  (\bibinfo{year}{2011}).

\bibitem{feldman96gene-culture}
\bibinfo{author}{Feldman, M.~W.} \& \bibinfo{author}{Laland, K.~N.}
\newblock \bibinfo{title}{Gene-culture coevolutionary theory}.
\newblock \emph{\bibinfo{journal}{Trends in Ecology \& Evolution}}
  \textbf{\bibinfo{volume}{11}}, \bibinfo{pages}{453--457}
  (\bibinfo{year}{1996}).

\bibitem{lehmann08the-coevolution}
\bibinfo{author}{Lehmann, L.} \& \bibinfo{author}{Feldman, M.~W.}
\newblock \bibinfo{title}{The coevolution of culturally transmitted and
  genetically encoded altruistic helping traits}.
\newblock \emph{\bibinfo{journal}{Theoretical Population Biology}}
  \textbf{\bibinfo{volume}{73}}, \bibinfo{pages}{506--516}
  (\bibinfo{year}{2008}).

\bibitem{ravigne09live}
\bibinfo{author}{Ravigne, V.}, \bibinfo{author}{Dieckmann, U.} \&
  \bibinfo{author}{Olivieri, I.}
\newblock \bibinfo{title}{Live where you thrive: joint evolution of habitat
  choice and local adaptation facilitates specialization and promotes
  diversity}.
\newblock \emph{\bibinfo{journal}{American Naturalist}}
  \textbf{\bibinfo{volume}{174}}, \bibinfo{pages}{E141--E169}
  (\bibinfo{year}{2009}).

\bibitem{laland99evolutionary}
\bibinfo{author}{Laland, K.~N.}, \bibinfo{author}{Odling-Smee, F.~J.} \&
  \bibinfo{author}{Feldman, M.~W.}
\newblock \bibinfo{title}{Evolutionary consequences of niche construction and
  their implications for ecology}.
\newblock \emph{\bibinfo{journal}{Proceedings of the National Academy of
  Sciences}} \textbf{\bibinfo{volume}{96}}, \bibinfo{pages}{10242--10247}
  (\bibinfo{year}{1999}).

\bibitem{odling-smee03niche}
\bibinfo{author}{Odling-Smee, F.~J.}, \bibinfo{author}{Laland, K.~N.} \&
  \bibinfo{author}{Feldman, M.~W.}
\newblock \emph{\bibinfo{title}{Niche Construction: The Neglected Process in
  Evolution}} (\bibinfo{publisher}{Princeton University Press},
  \bibinfo{address}{Princeton, NJ}, \bibinfo{year}{2003}).

\bibitem{west-eberhard03developmental}
\bibinfo{author}{West-Eberhard, M.~J.}
\newblock \emph{\bibinfo{title}{{Developmental Plasticity and Evolution}}}
  (\bibinfo{publisher}{Oxford University Press}, \bibinfo{address}{New York},
  \bibinfo{year}{2003}).

\bibitem{21phenotypic}
\bibinfo{editor}{Pfennig, D.~W.} (ed.) \emph{\bibinfo{title}{Phenotypic
  Plasticity \& Evolution: Causes, Consequences, Controversies}}
  (\bibinfo{publisher}{Taylor \& Francis}, \bibinfo{address}{Boca Raton,
  Florida}, \bibinfo{year}{2021}).

\bibitem{payne19the-causes}
\bibinfo{author}{Payne, J.~L.} \& \bibinfo{author}{Wagner, A.}
\newblock \bibinfo{title}{The causes of evolvability and their evolution}.
\newblock \emph{\bibinfo{journal}{Nature Reviews Genetics}}
  \textbf{\bibinfo{volume}{20}}, \bibinfo{pages}{24--38}
  (\bibinfo{year}{2019}).

\bibitem{wagner13robustness}
\bibinfo{author}{Wagner, A.}
\newblock \emph{\bibinfo{title}{Robustness and Evolvability in Living Systems}}
  (\bibinfo{publisher}{Princeton University Press},
  \bibinfo{address}{Princeton, NJ}, \bibinfo{year}{2013}).

\bibitem{earl04evolvability}
\bibinfo{author}{Earl, D.~J.} \& \bibinfo{author}{Deem, M.~W.}
\newblock \bibinfo{title}{Evolvability is a selectable trait}.
\newblock \emph{\bibinfo{journal}{Proceedings of the National Academy of
  Sciences}} \textbf{\bibinfo{volume}{101}}, \bibinfo{pages}{11531--11536}
  (\bibinfo{year}{2004}).

\bibitem{kirschner98evolvability}
\bibinfo{author}{Kirschner, M.} \& \bibinfo{author}{Gerhart, J.}
\newblock \bibinfo{title}{Evolvability}.
\newblock \emph{\bibinfo{journal}{Proceedings of the National Academy of
  Sciences USA}} \textbf{\bibinfo{volume}{95}}, \bibinfo{pages}{8420--8427}
  (\bibinfo{year}{1998}).

\bibitem{pigliucci08is-evolvability}
\bibinfo{author}{Pigliucci, M.}
\newblock \bibinfo{title}{Is evolvability evolvable?}
\newblock \emph{\bibinfo{journal}{Nature Reviews Genetics}}
  \textbf{\bibinfo{volume}{9}}, \bibinfo{pages}{75--82} (\bibinfo{year}{2008}).

\bibitem{wagner96complex}
\bibinfo{author}{Wagner, G.~P.} \& \bibinfo{author}{Altenberg, L.}
\newblock \bibinfo{title}{Complex adaptations and the evolution of
  evolvability}.
\newblock \emph{\bibinfo{journal}{Evolution}} \textbf{\bibinfo{volume}{50}},
  \bibinfo{pages}{967--976} (\bibinfo{year}{1996}).

\bibitem{otto97deleterious}
\bibinfo{author}{Otto, S.~P.} \& \bibinfo{author}{Feldman, M.~W.}
\newblock \bibinfo{title}{Deleterious mutations, variable epistatic
  interactions, and the evolution of recombination}.
\newblock \emph{\bibinfo{journal}{Theoretical Population Biology}}
  \textbf{\bibinfo{volume}{51}}, \bibinfo{pages}{134--147}
  (\bibinfo{year}{1997}).

\bibitem{otto02resolving}
\bibinfo{author}{Otto, S.~P.} \& \bibinfo{author}{Lenormand, T.}
\newblock \bibinfo{title}{Resolving the paradox of sex and recombination}.
\newblock \emph{\bibinfo{journal}{Nature Reviews Genetics}}
  \textbf{\bibinfo{volume}{3}}, \bibinfo{pages}{252--261}
  (\bibinfo{year}{2002}).

\bibitem{otto09the-evolutionary}
\bibinfo{author}{Otto, S.~P.}
\newblock \bibinfo{title}{The evolutionary enigma of sex}.
\newblock \emph{\bibinfo{journal}{American Naturalist}}
  \textbf{\bibinfo{volume}{174}}, \bibinfo{pages}{S1--S14}
  (\bibinfo{year}{2009}).

\end{thebibliography}
\end{document}